\documentclass[10pt, conference, letterpaper]{IEEEtran}

\IEEEoverridecommandlockouts

\usepackage[utf8]{inputenc}
\usepackage{bm}
\usepackage{xspace}

\usepackage{xcolor}
\newcommand{\nbo}[1]{{\sf\color{orange}[#1]}}
\newcommand{\ls}[1]{\nbo{LS: #1}}
\newcommand{\si}[1]{\textcolor{black}{#1}}
\newcommand{\yy}[1]{\textcolor{black}{#1}}
\newcommand{\com}[1]{{\sf\color{black}[#1]}}
\newcommand{\yl}[1]{\com{Yuezhou: #1}}
\newcommand{\blu}[1]{{\color{black}#1}}

\newcommand{\nodes}{\ensuremath{\mathcal{V}}\xspace}
\newcommand{\edges}{\ensuremath{\mathcal{E}}\xspace}
\newcommand{\sources}{\ensuremath{\mathcal{S}}\xspace}
\newcommand{\source}{\ensuremath{s}\xspace}
\newcommand{\learners}{\ensuremath{\mathcal{L}}\xspace}
\newcommand{\learner}{\ensuremath{\ell}\xspace}

\newcommand{\types}{\ensuremath{\mathcal{T}}\xspace}
\newcommand{\type}{\ensuremath{t}\xspace}
\newcommand{\xset}{\ensuremath{\mathcal{X}}\xspace}

\newcommand{\nsamp}{\ensuremath{N}}
\newcommand{\inc}{\mathtt{in}}
\newcommand{\out}{\mathtt{out}}
\newcommand{\feasibleset}{\mathcal{D}}

\newcommand{\Count}{N}
\newcommand{\x}{\vc{x}}

\usepackage{amsthm}

\newtheorem{lem}{\textbf{Lemma}}
\newtheorem{thm}{\textbf{Theorem}}
\newtheorem{Def}{\textbf{Definition}}

\newtheorem{ass}{\textbf{Assumption}}

\newcommand{\expect}[1]{\mathbb{E}\left[ #1 \right]}
\usepackage[ruled,vlined]{algorithm2e}
\newcommand{\reals}{\mathbb{R}}
\newcommand{\naturals}{\mathbb{N}}
\newcommand{\cov}{\mathtt{cov}}
\newcommand{\map}{\mathtt{MAP}}

\usepackage{amsmath,amssymb,amsfonts}
\usepackage{algorithmic}

\usepackage{graphicx}
\usepackage{verbatim}
\usepackage{subfigure}
\usepackage{float}
\usepackage{textcomp}
\usepackage{xcolor}
\usepackage{mathtools}
\usepackage{cite}
\usepackage{paralist}
\usepackage{url}
\usepackage{hyperref}
\usepackage{authblk}
\usepackage{fancyhdr}
\usepackage{multicol}
\usepackage{multirow}

\newcommand{\vc}[1]{\boldsymbol{#1}}

\newcommand{\arxiv}[2]{#1} 

\title{Experimental Design Networks: \\
A Paradigm for Serving Heterogeneous  Learners under Networking Constraints
\thanks{* Y. Liu and Y. Li contributed equally to the paper.}
}
\author[ ]{Yuezhou Liu*, Yuanyuan Li*, Lili Su, Edmund Yeh, Stratis Ioannidis}
\affil[ ]{Department of Electrical and Computer Engineering\\ Northeastern University, Boston, MA, USA}
\affil[ ]{\textit{liu.yuez@northeastern.edu, yuanyuanli@ece.neu.edu, l.su@northeastern.edu, \{eyeh, ioannidis\}@ece.neu.edu}}

\begin{document}

\maketitle

\thispagestyle{fancy}
\lhead{} 
\chead{}
\rhead{} 
\lfoot{} 
\cfoot{\thepage} 
\renewcommand{\headrulewidth}{0pt}
\renewcommand{\footrulewidth}{0pt} 
\pagestyle{fancy}

\begin{abstract}
Significant advances in edge computing capabilities enable learning to occur at geographically diverse locations. In general, the training data needed in those learning tasks are not only heterogeneous but also not fully generated locally.  
In this paper, we propose an {\em experimental design network} paradigm, 
wherein learner nodes train possibly different Bayesian linear regression models via consuming data streams generated by data source nodes over a network. 
We formulate this problem as a social welfare optimization problem in which the global objective is defined as 
the sum of experimental design objectives of individual learners, and the decision variables are the data transmission strategies subject to network constraints. 
We first show that, assuming Poisson data streams, 
the global objective is a continuous DR-submodular function. We then propose a Frank-Wolfe type algorithm that outputs a solution within a $1-1/e$ factor from the optimal. Our algorithm contains 
a novel gradient estimation component which is carefully designed based on Poisson tail bounds and sampling.  
Finally, we complement our theoretical findings through extensive experiments.
Our numerical evaluation shows that the proposed algorithm outperforms several baseline algorithms both in maximizing the global objective and in the quality of the trained models. 

\end{abstract}



\section{Introduction}\label{sec: intro}

We study a network in which heterogeneous learners dispersed at different locations perform  local learning tasks by  fetching relevant yet remote data. 
Concretely, data sources generate data streams containing  both features and labels/responses, which are transmitted over the network (potentially through several intermediate router nodes) towards  learner nodes. Generated data samples are   used by learners to train models locally. 
We are interested in the design of rate allocation strategies that maximize the model training quality of learner nodes, subject to network constraints.  
This problem is relevant in practice.  For example, in a mobile edge computing network \cite{abbas2017mobile,yang2019mobile}, data are generated by  end devices such as mobile phones (data sources) and sent to edge servers (learners) for model training, a relatively intensive computation. 
In a smart city \cite{albino2015smart,mohammadi2018enabling}, we can collect various types of data such as image, temperature, humidity, traffic, and seismic measurements, from different sensors. These data could be used to forecast transportation traffic,  the spread of disease, pollution levels, the weather,  and so on, while training for each task could happen at different public service entities. 

We quantify the impact that data samples have on learner model training accuracy  by leveraging objectives motivated by  
\emph{experimental design} \cite{boyd2004convex}, a classic problem in statistics and machine learning. 
\blu{This problem 
arises in many machine learning and data mining settings, including recommender systems \cite{deshpande2012linear}, active learning~\cite{settles2009active}, and data preparation~\cite{polyzotis2018data}, to name a few.}
In standard experimental design,  a learner decides on 
which experiments to conduct 
\blu{so that, under budget constraints, an 
 objective modeling prediction accuracy is maximized. }
 Learner objectives are usually scalarizations of the estimation error covariance. 

In this paper, 
we propose \emph{experimental design networks}, \blu{a novel optimization framework
that extends classic experimental problems to}
maximize 
the sum of experimental design objectives across 
networked  learners. 
Assuming Poisson data streams \blu{and Bayesian linear  regression as the learning task}, we define the utility of a learner as the expectation of its so-called D-optimal design objective~\cite{boyd2004convex}\yy{, namely, the log-determinant of the learner's estimation error covariance matrix}. 
Our goal is to determine the data rate allocation of   each  network edge that maximizes the aggregate utility across learners. 
Extending experimental design to networked learners is 
non-trivial. Literature on experimental design for machine learning considers budgets \si{imposed} on the number 
of data samples used to train the model \cite{horel2014budget,guo2019accelerated,gast2020linear, guo2018experimental,flaherty2006robust}. Instead, we consider far more complex constraints on the data transmission rates across the network, as determined by network link capacities, the network topology, and  data generation rates at sources. 



To the best of our knowledge, we are the first to study such a networked learning problem, wherein  learning tasks at  heterogeneous learners are coupled via data transmission constraints over an arbitrary network topology. 
Our detailed contributions are as follows:
\begin{itemize}
    \item We \yy{are the first to introduce and formalise} the experimental design network problem, which enables the study of multi-hop data transmission strategies for distributed learning over arbitrary network topologies.
    \item We prove that, \blu{assuming Poisson data streams, Bayesian linear regression as the learning task,} and D-optimal design objectives at the learners, our framework leads to the maximization of continuous DR-submodular objective subject to a lower-bounded convex constraint set.
    \item Though the objective is not concave, we propose a polynomial-time algorithm based on a variant of the Frank-Wolfe algorithm \cite{bian2017guaranteed}. \si{To do so, we introduce and analyse a novel gradient estimation procedure, tailored to  Poisson  data streams}. 
    We show that the proposed algorithm, \si{coupled with  our novel gradient estimation,} is guaranteed to produce a solution within a $1-1/e$ approximation factor from the optimal.
    \item We conduct extensive evaluations over 
    different network topologies, showing that our proposed algorithm outperforms several baselines in both maximizing the objective function and in the quality of trained target models. 
\end{itemize}

The rest of this paper is organized as follows. In Sections \ref{sec: related} and \ref{sec: pre}, we review related work and provide technical preliminaries. Section \ref{sec: problem} introduces  our framework of experimental design networks. Section \ref{sec: result} describes our proposed algorithm. We discuss extensions in Section \ref{sec: extension} and present numerical experiments in Section \ref{sec: simulation}. 
We conclude in Section \ref{sec: conclusion}.

\section{Related Work}\label{sec: related}

\noindent\textbf{Distributed Computing/Learning in Networks.}
Distribution of computation tasks has been studied in hierarchical edge cloud networks~\cite{tong2016hierarchical}, multi-cell mobile networks~\cite{poularakis2019joint}, and joint with data caching in arbitrary  networks~\cite{kamran2019deco}. 
There is a rich literature on distributing machine learning computations over networks, including exchanging gradients in federated learning~\cite{lalitha2019peer,neglia2019role,wang2018edge}, states in reinforcement learning~\cite{zhang2018fully}, and data vs.~model offloading~\cite{wang2021network} among collaborating neighbor nodes. We depart from the aforementioned works in  (a) considering multiple learners with distinct learning tasks, (b) introducing experimental design objectives, quite  different from objectives considered above,  (c) studying a multi-hop network, and (d) focusing on the optimization of streaming data movements, as opposed to gradients or intermediate result computations.

\vskip 0.2\baselineskip 
\noindent\textbf{Experimental Design.} The experimental design problem is classic and well-studied \cite{boyd2004convex, pukelsheim2006optimal}. 
Several works study the 
D-optimality objective \cite{horel2014budget,guo2019accelerated,gast2020linear, guo2018experimental,huan2013simulation} for a single learner subject to budget constraints on the cost for conducting the experiments.
Departing from  previous work, we study a problem involving multiple learners subject to more complex constraints, induced by the network. Our problem also falls in the \emph{continuous} DR-submodular setting, departing from the discrete setting in prior work. In fact, our work is the first to show that such an optimization, with Poisson data streams,  
can be solved via continuous DR-submodularity techniques. 

\vskip 0.2\baselineskip 
\noindent\textbf{DR-submodular Optimization.} 
Submodularity is traditionally studied in the context of set functions \cite{nemhauser1978analysis,calinescu2011maximizing}, but was recently extended to functions over the integer lattice \cite{soma2015generalization} and the continuous domain \cite{bian2017guaranteed}. 
Despite the non-convexity and the general NP-hardness of the problem, when the constraint set is down-closed and convex, maximizing monotone continuous DR-submodular functions 
can be done in polynomial time via a variant of the Frank-Wolfe algorithm.  This yields a solution within $1-1/e$ from the optimal~\cite{bian2017guaranteed,calinescu2011maximizing}, outperforming the projected gradient ascent method, which provides $1/2$ approximation guarantee over arbitrary convex constraints~\cite{hassani2017gradient}. 

The  continuous greedy algorithm \cite{calinescu2011maximizing} maximizes a submodular set function subject to matroid constraints: this first applies the aforementioned Frank-Wolfe variant to the so-called \emph{multilinear relaxation} of the discrete submodular function, and subsequently uses rounding \cite{ageev2004pipage,chekuri2010dependent}. The multilinear relaxation of a submodular function is in fact a canonical example of a continuous DR-submodular function,  whose optimization comes with the aforementioned guarantees.
Our objective function results from a \emph{new continuous relaxation}, which we introduce in this paper for the first time.  In particular, we show that assuming a Poisson distribution on inputs on the (integer lattice) DR-submodular function of D-optimal design  yields a continuous DR-submodular function. This ``Poisson'' relaxation is directly motivated by our networking problem,  is distinct from the multilinear relaxation  \cite{soma2018maximizing,calinescu2011maximizing,hassani2017gradient}, and requires a \si{novel} gradient estimation procedure. \si{Our} constraint set \si{also} requires special treatment as it is not down-closed\yy{, as required by the aforementioned Frank-Wolfe variant~\cite{bian2017guaranteed,calinescu2011maximizing}}; nevertheless, we   attain a $1-1/e$ \si{approximation, improving upon} the \yy{$1/2$ factor of}  projected gradient ascent~\cite{hassani2017gradient}.


\vskip 0.2\baselineskip 
\noindent\textbf{Submodularity in Networking and Learning.}
Submodular functions are widely encountered in studies of \si{both} networking and machine learning. 
Submodular objectives 
appear in studies of network caching \cite{ioannidis2018adaptive, poularakis2016complexity}, routing \cite{ioannidis2018jointly}, rate allocation \cite{kamran2021rate}, sensor network design \cite{wu2019charging}, as well as placement of virtual network functions \cite{sallam2019joint}. 
Submodular utilities 
are used for data collection in sensor networks \cite{zheng2014submodular} and also the design of incentive mechanisms for mobile phone sensing \cite{yang2012crowdsourcing}. Many machine learning problems are submodular, including structure learning, clustering, feature selection, and active learning~(see e.g., \cite{krause2008beyond}). Our proposed experimental design network paradigm
expands this list in a novel way.

\section{Technical Preliminary}\label{sec: pre}
%
%

We begin with a technical preliminary on linear regression, experimental design, and DR-submodularity. The contents of this section are classic; for additional details, we refer the interested reader to, e.g., \cite{james2013introduction,gallager2013stochastic} for linear regression, \cite{boyd2004convex} for experimental design, and \cite{bian2017guaranteed} for submodularity.

\subsection{Bayesian Linear Regression }
\label{sec: LR}
%

In the standard linear regression setting, a learner observes $n$ samples $(\vc{x}_i,y_i)$, $i=1,\ldots,n$, where $\vc{x}_i\in \reals^d$ and $y_i\in \reals$ are the feature vector and label of sample $i$, respectively. Labels are assumed to be linearly related to the features; in particular, there exists a model parameter vector $\vc{\beta}\in \reals^d$ such that 
\begin{align}
    y_i = \vc{x}_i^\top\vc{\beta} + \epsilon_i,\quad \text{for all}~ i\in\{1,\dots,n\}, \label{eq:linear}
\end{align}
and $\epsilon_i$ are i.i.d. zero mean normal noise variables with variance $\sigma^2$ (i.e., $\epsilon_i\sim N(0, \sigma^2)$). 

The learner's goal is to estimate the model parameter $\vc{\beta}$ from samples $\{(\vc{x_i},y_i)\}_{i=1}^n$. In Bayesian linear regression, it is additionaly assumed that $\vc{\beta}$ is sampled from a prior normal distribution with  mean $\vc{\beta}_0\in\reals^d$ and covariance $\vc{\Sigma}_0\in\reals^{d\times d}$ (i.e., $\vc{\beta} \sim N(\vc{\beta}_0, \vc{\Sigma}_0)$). Under this prior, given dataset $\{(\vc{x_i},y_i)\}_{i=1}^n$, maximum a posteriori (MAP) estimation of $\vc{\beta}$ amounts to \cite{gallager2013stochastic}:
\begin{align}
   \begin{split} \hat{\vc{\beta}}_\map 
    &= (\vc{X}^\top \vc{X} + \sigma^2\vc{\Sigma}_0^{-1})^{-1} \vc{X}^{\top}\vc{y}\\&\quad + (\vc{X}^\top \vc{X} + \sigma^2\vc{\Sigma}_0^{-1})^{-1}\sigma^2\vc{\Sigma}_0^{-1}\vc{\beta}_0, \end{split}\label{eq:map}
\end{align}
where $\vc{X} = [\vc{x}_i^\top]_{i=1}^n\in\reals^{n\times d}$ is the matrix of features, $\vc{y}\in\reals^n$ is the vector of labels,  $\sigma^2$ is the noise variance, and $\vc{\beta}_0,\vc{\Sigma}_0$ are the mean and covariance of the prior, respectively. We note that,  in practice, the inherent noise variance $\sigma^2$ is often not known, and is typically treated as a regularization parameter and determined via cross-validation. 

The quality of this estimator can be characterized by the covariance of the estimation error difference $\hat{\vc{\beta}}_\map-\vc{\beta}$ (see, e.g., Eq. (10.55) in \cite{gallager2013stochastic}):
\begin{equation}
    \cov(\hat{\vc{\beta}}_\map-\vc{\beta}) = \big(\frac{1}{\sigma^2}\vc{X}^{\mathrm{T}}\vc{X} + \vc{\Sigma}_0^{-1}\big)^{-1} \in \reals^{d\times d}. \label{eq:cov} 
\end{equation}
The covariance summarizes  estimator quality in all directions in $\reals^d$:  given an unseen sample $(\vc{x},y)\in\reals^d\times \reals$, also obeying~\eqref{eq:linear}, the expected prediction error (EPE) is given by
\begin{align}\label{eq:epe}
    \expect{(y-\vc{x}^\top \hat{\vc{\beta}}_\map)^2} = \sigma^2 + \vc{x}^\top  \cov(\hat{\vc{\beta}}_\map-\vc{\beta})\vc{x}.
\end{align}
Hence, the eigenvalues of Eq.~\eqref{eq:cov}
 capture the overall variability of the expected prediction error in different directions.
 
 \subsection{Experimental Design}
 \label{sec: experimental design for LR}
 
 In experimental design, a learner 
determines which experiments to conduct to learn the most accurate linear model. Formally, given $p$ possible experiment settings, each described by feature vectors $\vc{x}_i\in\reals^d$, $i=1,\ldots,p$, the learner selects a total of $n$  experiments to conduct with these feature vectors, possibly with repetitions,\footnote{Note that, due to the presence of noise in labels, repeating the same experiment makes intuitive sense; formally, repetition of an experiment with features $\vc{x}_i$ reduces the EPE \eqref{eq:epe} in this direction.} collects associated labels, and then performs linear regression on these sample pairs. In classic experimental design (see, e.g., \cite{boyd2004convex}), the selection is formulated as an optimization problem minimizing a scalarization of the covariance \eqref{eq:cov}. For example, in D-optimal design, the vector $\vc{n}=[n_i]_{i=1}^p\in \mathbb{N}^p$ of the number of times each experiment is to be performed is determined by minimizing 
$$\log\det[ \cov(\hat{\vc{\beta}}_\map-\vc{\beta})]\stackrel{\eqref{eq:cov}}{=} \log\det\big[\big(\sum_{i=1}^p \frac{n_i}{\sigma^2} \vc{x}_i\vc{x}_i^\top + \vc{\Sigma}_0^{-1}\big)^{-1}\big] $$
or, equivalently, by solving the  maximization problem:
\begin{subequations}
\label{eq:doptprob}
\begin{align}
    \text{Max.:}~~&G(\vc{n};\sigma,\!\vc{\Sigma}_0) \equiv \log\det\big(\textstyle\sum_{i=1}^p \!\frac{n_i}{\sigma^2}\vc{x}_i\vc{x}_i^{\top} + \vc{\Sigma}_0^{-1}\big),\label{eq: D-optimal} \\
    \text{s.t.:}~~&\textstyle\sum_{i=1}^p n_i = n.
\end{align}
\end{subequations}
In other words, $\vc{n}\in \naturals^p$ is selected in such a way so that the $\log\det[ \cov(\hat{\vc{\beta}}_\map-\vc{\beta})]$ is as small as possible. Intuitively, this amounts to selecting the experiments that minimize the product of the eigenvalues of the covariance;\footnote{Other commonly encountered scalarizations~\cite{boyd2004convex} behave similarly. E.g., E-optimality minimizes the maximum eigenvalue, while A-optimality minimizes the sum of the eigenvalues.} alternatively, Prob.~\eqref{eq:doptprob} also  maximizes the mutual information between the  labels $\vc{y}$ (to be collected) and $\hat{\vc{\beta}}_\map$; in both interpretations, the selection aims to pick experiments in a way that minimizes the variability of the resulting estimator $\hat{\vc{\beta}}_\map$.

\subsection{DR-Submodularity}\label{sec: submodular}
We introduce here  diminishing-returns submodularity:
\begin{Def}[DR-Submodularity \cite{bian2017guaranteed,soma2015generalization}]\label{def: submodular}
A function $f: \naturals^p \to \reals$ is called \emph{diminishing-returns (DR) submodular} iff for all $\vc{x}, \vc{y} \in \naturals^p$ such that $\vc{x}\leq \vc{y}$ and all $k\in \naturals$, 
\begin{equation}\label{eq:drsub}
    f(\vc{x} \!+\! k \vc{e}_j)\! - \! f(\vc{x}) \!\geq\! f(\vc{y} \!+\! k\vc{e}_j)\! -\! f(\vc{y}),~\text{for all }j=1,\ldots,p,
\end{equation}
where $\vc{e}_j$ is the $j$-th  standard basis vector. 

Moreover, if \eqref{eq:drsub} holds for a real valued function $f: \reals^p_+ \to \reals$ for all  $\vc{x}, \vc{y} \in \reals^p$ such that $\vc{x}\leq \vc{y}$ and all $k\in \reals_+$, the function is called \emph{continuous DR-submodular}. 
\end{Def}
The above definition generalizes the submodularity of set functions (whose domain is $\{0,1\}^p$) to  functions over integer lattice (in the case of DR-submodularity), and continuous functions (in the case of continuous DR-submodularity). 
%
Particularly for continuous functions, if $f$ is differentiable, continuous DR-submodularity is equivalent to $\nabla f$ being antitone. Moreover, if $f$ is twice-differentialble, $f$ is continuous DR-submodular if all entries of its Hessian $\nabla^2 f$ are non-positive. 
%
%
%
 DR-submodularity is directly pertinent to D-optimal design:
\begin{lem}[Horel et al.~\cite{horel2014budget}]\label{lem: supermodularity}
Function $G: \mathbb{N}^p \to\mathbb{R}_+$ in \eqref{eq: D-optimal} is (a) monotone-increasing and (b) DR-submodular. 
%
\end{lem}
For completeness, we provide a proof in Appendix~\ref{proof: supermodular}.
An immediate consequence of this lemma is that polynomial-time approximation algorithms exist to solve Prob.~\eqref{eq:doptprob} \si{with a $1-1/e$  guarantee} (see, e.g.,~\cite{soma2018maximizing,horel2014budget}), although \si{Prob.~\eqref{eq:doptprob}} is a classic NP-hard problem \cite{horel2014budget}. 

\section{Problem Formulation}
\label{sec: problem}
We consider a network that aims to facilitate a distributed learning task.   The network comprises (a) data source nodes (e.g., sensors, test sites, experimental facilities, etc.) that generate streams of data, (b) learner nodes, that consume data with the purpose of training  models,  and (c) intermediate nodes (e.g., routers), that facilitate the communication of data from sources to learners. The data that learners wish to consume is determined by experimental design objectives, akin to the ones described in Sec.~\ref{sec: experimental design for LR}. Our goal is to design network communications in an optimal fashion, that maximizes learner social welfare. We describe each of the above system components in more detail below.

\subsection{Network Model.}\label{sec: system model} We model the above system as general multi-hop network with a topology represented by a directed acyclic graph (DAG) $\mathcal{G}(\nodes, \edges)$, where $\nodes$ is the set of nodes and $\edges\subset \nodes\times\nodes$ is the set of links.  Each link $e=(u,v)\in\edges$ can transmit data at a maximum rate (i.e., link capacity) $\mu^e\geq 0$. Sources $\sources\subset \nodes $ of the DAG 
(i.e., nodes with no incoming edges)
generate data streams, while learners $\learners\subset \nodes$ reside at DAG sinks
(nodes with no outgoing edges). \si{We assume this for simplicity; we discuss how to remove this assumption, and how to generalize our analysis beyond DAGs, in Sec.~\ref{sec: extension}.}

\noindent\textbf{Data Sources.}
 Each data source $\source\in\sources$ generates a stream of labeled data. In particular, we assume that there exists a finite\footnote{We extend this to a setting where experiments  are infinite in Sec.~\ref{sec: extension}.} set $\xset\subset\reals^d$ of experiments every source can conduct. Once experiment with features $\vc{x}\in \xset$ is conducted, the source can label it with a label $y \in \reals$ of type $\type$ out of a set of possible types $\types$. Intuitively, features $\vc{x}$ correspond to parameters set in an experiment (e.g., pixel values in an image, etc.),  label types $\type\in\types$ correspond to possible measurements (e.g., temperature, radiation level, etc.), and labels $y$ correspond to the actual measurement value collected (e.g., $23^{\circ}\text{C})$. 

We assume that every source generates labeled pairs $(\vc{x},y)\in \reals^d\times \reals$ of type $\type$ according to a Poisson process of rate $\lambda_{\x,\type}^{\source}\geq 0.$ Moreover, we assume that generated data follows a linear model \eqref{eq:linear}; that is, for every type $\type\in\types$, there exists a $\vc{\beta}_\type\in\reals^d$ s.t.
 $y = \vc{x}^\top\vc{\beta}_\type + \epsilon_\type$
where $\epsilon_\type\in \reals$ are i.i.d.~zero mean normal noise variables with variance $\sigma^2_\type>0$, independent across experiments and sources $\source\in\sources$.

\noindent\textbf{Learners}. 
Each learner $\learner \in \learners$ wishes to learn a model $\vc{\beta}_{\type^\learner}$ for some type $\type^\learner\in \types$. We assume that each learner has a prior $N(\vc{\beta}_\learner,\vc{\Sigma}_\learner)$ on $\vc{\beta}_{\type^\learner}$. The learner wishes to use the network to receive data pairs $(\x,y)$ of type $\type^\learner$, and subsequently estimate $\vc{\beta}_{\type^\learner}$
through the MAP estimator \eqref{eq:map}.
Note that two learners $\learner$, $\learner'$ may be interested to learn the same model (if $\type^\learner=\type^{\learner'}$).

\noindent\textbf{Network Constraints}. The different data pairs $(\x,y)\in\reals^d\times\reals$ generated by sources are transmitted over edges in the network along with their types $\type \in \types$ and eventually delivered to learners. Our network design aims at allocating network capacity  to different flows to meet learner needs.\footnote{We  assume  hop-by-hop routing; see Sec.~\ref{sec: extension} for an extension to source routing.} For each edge $e\in\edges$, we denote  the rate with which data pairs of type $\type\in\types$ with features $\vc{x}\in\xset$ are transmitted as $\lambda_{\vc{x},\type}^e\geq 0.$  
We also denote by
\begin{align}\label{eq: incoming flow equation}
    \lambda_{\vc{x},\type}^{v,\inc} \equiv \begin{cases} 
    \lambda_{{\vc{x},\type}}^v, &\text{if~}v\in\sources,\\ \sum_{(u,v)\in \edges}\lambda_{\vc{x},\type}^{(u,v)}, & \text{o.w.}\end{cases} 
\end{align}
the corresponding incoming traffic to node $v\in \nodes$, and
\begin{align}
    \lambda_{\vc{x},\type}^{v,\out} = \sum_{(v,u)\in \edges}\lambda_{\vc{x},\type}^{(v,u)}\label{eq: lin}
\end{align}
the corresponding outgoing traffic from $v\in\nodes$.
Specifically for learners, we denote by
\begin{align}\label{eq: learner flow}
    \lambda^\learner_{\vc{x}} \equiv \lambda^{\learner,\inc}_{\vc{x},\type^\learner},~\text{and}~\vc{\lambda}^{\learner}=[\lambda^\learner_{\vc{x}}]_{\vc{x}\in \xset}\in \reals^{|\xset|},~~\text{for all}~\learner\in \learners,
\end{align}
the incoming traffic with different features $\vc{x}\in \xset$ of type $\type^\learner$ at  $\learner\in\learners$.
To satisfy capacity constraints, we must have
\begin{align}\sum_{\vc{x}\in \mathcal{X}, \type\in \types} \lambda_{\vc{x},\type}^e\leq \mu^e \label{eq: cap}, \quad\text{for all }e\in\edges,\end{align} 
while flow bounds imply that
\begin{align} \lambda_{\vc{x},\type}^{v,\out} \leq  \lambda_{\vc{x},\type}^{v,\inc}, \quad\text{for all }\vc{x}\in\xset,\type \in\types, v\in \nodes\setminus \learners,  \label{eq: flow}\end{align}
as data pairs can be dropped. We denote by
\begin{align}\vc{\lambda} = \big[[\lambda_{\vc{x},\type}^e]_{\vc{x}\in\xset,\type\in\types,e\in\edges}; [\lambda_{\vc{x}}^\learner]_{\vc{x}\in\xset,\learner\in\learners}\big]\end{align}  the vector comprising edge and learner rates. Let
\begin{align}\feasibleset = \big\{\vc{\lambda}\in \reals_+^{|\xset||\types| |\edges|}\!\!\times \reals^{|\xset||\learners|}_+\text{ that satisfy \eqref{eq: learner flow}--\eqref{eq: flow}}\big\},\! \end{align}
be the feasible set of edge rates and learner rates. We make following assumption on the network substrate:
\begin{ass}\label{asm:pois}
For $\vc{\lambda}\in\feasibleset$, the system is stable and, in steady state, pairs $(\x,y)\in\reals^{d}\times \reals$ of type $t^\learner$ arrive at learner $\learner\in \learners$ according to $|\xset|$ independent Poisson processes with rate $\lambda^\learner_{\x}$.
\end{ass}
This is satisfied, if, e.g., the network is a Kelly network \cite{kelly2011reversibility} of $M/M/1$ queues, $M/M/c$ queues, etc., under FIFO, Last-In First-Out (LIFO), and processor sharing service disciplines, or other queues for which Burke's theorem holds \cite{gallager2013stochastic}. 

\subsection{Networked Learning Problem}
\label{sec: networked learning}
We consider a data acquisition time period $T$, at the end of which 
each learner $\learner\in\learners$ estimates $\vc{\beta}_{\type^\learner}$ 
based on the data it has received during this period 
via MAP estimation.
Under Assumption~\ref{asm:pois},  the arrivals of pertinent data pairs at learner $\learner$ \si{form a} Poisson process with rate $\lambda_{\x}^\learner$. Let $n_{\x}^\learner\in\naturals$ be the cumulative number of times that a pair $(\x,y)$ of type $\type^\learner$ was collected by learner $\learner$ during this period, and $\vc{n}^\learner = [n^{\learner}_{\vc{x}}]_{\vc{x}\in \xset}$ the vector of arrivals across all experiments. Then, 
\begin{align}\label{eq:prod}
    \mathrm{Pr}[\vc{n}^\learner=\vc{n}] = \prod_{\vc{x}\in\xset}\frac{(\lambda_{\vc{x}}^\learner T)^{n_{\vc{x}}}  e^{-\lambda_{\vc{x}}^\learner T}}{n_{\vc{x}}!}, 
\end{align}
for all $\vc{n}=[n_{\vc{x}}]_{\vc{x}\in \xset}\in\naturals^{|\xset|}$ and $\learner\in\learners$.
Motivated by standard experimental design (see Sec.~\ref{sec: experimental design for LR}), 
we define the utility at learner $\learner\in\learners$ as the following expectation:
\begin{align}\label{eq: utility for single learner}
\begin{split}
    U^\learner (\vc{\lambda}^\learner) &= \mathbb{E}_{\lambda^\learner}\big[ G^\learner(\vc{n}^\learner)\big]\\&=
    \sum_{\vc{n}\in\naturals^{|\xset|}}G^\learner(\vc{n})\cdot\! \mathrm{Pr}[\vc{n}^\learner=\vc{n}],
    \end{split}
\end{align}
where 
$G^\learner(\vc{n}^\learner) \equiv G(\vc{n}^\learner;\sigma_{\type^\learner},\vc{\Sigma}_\learner)$ and $G$ is given by \eqref{eq: D-optimal}. 
We wish to solve the following  problem:
\begin{subequations}
\label{prob: utility}
\begin{align}
    \text{Maximize:} \quad &U(\vc{\lambda}) 
    =\sum_{\learner\in\learners}(U^\learner(\vc{\lambda}^\learner) - U^\learner(\mathbf{0})),\label{eq: objective function}\\
    \text{s.t.} \quad &\vc{\lambda}\in \feasibleset. 
\end{align}
\end{subequations}   
Indexing flows by both type $\type$ and features $\x$ implies that, to implement a solution $\vc{\lambda}\in \feasibleset$, routing decisions at intermediate nodes should be based on both quantities. Problem~\eqref{prob: utility} is non-convex in general.\footnote{It is easy to construct instances of objective \eqref{prob: utility} that are non-concave. For example, when $|\learners| = 1$, $d=1$, $\xset = \{0.1618, 0.3116\}$, $\sigma = 0.0422$, and $\Sigma_\learner = 0.2962$, the Hessian matrix is not negative semi-definite.} \si{Nevertheless,} we construct a polynomial time approximation algorithm in the next section.

\section{Main Results}\label{sec: result}
Our main contribution is to show that there exists a polynomial-time randomized algorithm that solves Prob.~\eqref{prob: utility} within a $1-1/e$ approximation ratio. We do so by establishing that the objective function in Eq.~\eqref{eq: objective function} is \emph{continuous DR- submodular} (see Definition~\ref{def:  submodular}). 

\subsection{Continuous DR-submodularity}

Our first main result establishes the continuous DR-submodularity of  the objective \eqref{eq: objective function}:
\begin{thm}\label{Thm: dr-submodular}
The objective function $U(\vc{\lambda})$ given by \eqref{eq: objective function} is monotone increasing and continuous DR-submodular in $\vc{\lambda}\in \reals_+^{|\xset|\times|\types|\times |\edges|}$.  
Moreover, 
\begin{align}\label{eq: partial derivative}
\frac{\partial U}{\partial\lambda^\learner_{\vc{x}}}  = T \sum_{n=0}^\infty\Delta^\learner_{\vc{x}}(\vc{\lambda}^\learner,n)\mathrm{Pr}[n^\learner_{\vc{x}} = n], 
\end{align}
where $\vc{n}^\learner$ is distributed as in Eq.~\eqref{eq:prod} and
\si{$\Delta^\learner_{\vc{x}}(\vc{\lambda}^\learner,n)$ is: } $$\mathbb{E}\left[G^\learner(\vc{n}^\learner)|n^\learner_{\vc{x}}= n+1\right] - \mathbb{E}\left[G^\learner(\vc{n}^\learner)|n^\learner_{\vc{x}}= n\right]> 0.$$ 
\end{thm}
The proof can be found in Section \ref{proof dr-submodular}; we establish the positivity of the gradient and non-positivity of the Hessian of $U$. We note that Theorem~\ref{Thm: dr-submodular} identifies a \emph{new type of continuous relaxation to DR-submodular functions}, via Poisson sampling; this is in contrast to the multilinear relaxation \cite{soma2018maximizing,calinescu2011maximizing,hassani2017gradient}, which is  ubiquitous in the literature \si{and relies on  Bernoulli sampling}.  
\yy{Finally, though our objective  is monotone and continuous DR-submodular, the constraint \si{set $\feasibleset$} is \emph{not} down-closed. Hence, the analysis by Bian et al.~\cite{bian2017guaranteed} does not directly apply, while using projected gradient ascent~\cite{hassani2017gradient} would only yield a $1/2$ approximation guarantee. }

\subsection{Algorithm and Theoretical Guarantee}
\label{sec: alg}
Our algorithm is summarized in Algorithm \ref{alg: F-W}. 
We follow the Frank-Wolfe variant for monotone DR-submodular function maximization by Bian et al.~\cite{bian2017guaranteed}, deviating both in the nature of the constraint set $\feasibleset$, but also, most importantly, in the way we estimate the gradients of objective $U$. 

%
\begin{algorithm}[t]
\label{alg: F-W}
\caption{Frank-Wolfe Variant}
\LinesNumbered
\KwIn{ $U:\feasibleset\to\reals_+$,  $\feasibleset$, step-size $\delta\in(0,1]$.}
$\lambda^0=0, \tau=0, k=0$ \\
\While{$\tau<1$}{
	find $\vc{v}^k$ s.t. $  \vc{v}^k = \mathop{\arg\max}_{\vc{v}\in \mathcal{D }}\langle \vc{v},\widehat{\nabla U(\vc{\lambda}^k)}\rangle$ \\
  $\gamma_k = \min \{ \delta, 1-\tau \}$ \\
  $\vc{\lambda}^{k+1} = \vc{\lambda}^k + \gamma_k \vc{v}^k, \tau=\tau+\gamma_k, k=k+1$
}
\Return $\lambda^K$
\end{algorithm}

%
\noindent\textbf{Frank-Wolfe Variant.} In the proposed Frank-Wolfe variant,  variables $\vc{\lambda}^k$ and $\vc{v}^k$ denote the solution and update direction at the $k$-th iteration, respectively. Starting from $\vc{\lambda}^0=\mathbf{0}\in \feasibleset$, the algorithm iterates as follows:
\begin{subequations}
\begin{align}
	& \vc{v}^k = \mathop{\arg\max}_{\vc{v}\in \mathcal{D }}\langle \vc{v},\widehat{\nabla U(\vc{\lambda}^k)}\rangle, \label{eq: vk update}\\
    & \vc{\lambda}^{k+1} = \vc{\lambda}^k + \gamma_k \vc{v}^k,\label{eq: lambda k update}
\end{align}
\end{subequations}
where $\gamma^k\in (0,1]$ is the stepsize with which we move along direction $\vc{v}^k$, and $\widehat{\nabla U(\cdot)}$ is an estimator of the gradient $\nabla U$ w.r.t. $[\lambda_{\vc{x}}^\learner]_{\vc{x}\in\xset,\learner\in\learners}$. The step size is set to $\delta>0$ for all but the last step, where it is selected so that the total sum of step sizes equals 1.  

We note that we face two challenges preventing us from computing the gradient of $\nabla U$ directly via. Eq.~\eqref{eq: partial derivative}: (a) the gradient computation involves an infinite summation over $n\in \naturals$, 
   and (b) conditional expectations in  $\Delta^\learner_{\vc{x}}(\vc{\lambda}^\learner,n)$  require further computing $|\xset|-1$ infinite sums.
Using \eqref{eq: partial derivative} directly in Algorithm~\ref{alg: F-W} would thus  not yield a polynomial-time algorithm.
To that end, we replace the gradient $\nabla U(\vc{\lambda}^k)$ used in the standard Frank-Wolfe method by an estimator, which we describe next. 

\noindent\textbf{Gradient Estimator.} Our estimator addresses challenge (a) above by truncating the infinite sum, and (b) via sampling. In particular,
for $n'\geq \lambda^\learner_{\x} T$, we estimate partial derivatives via the partial summation:
\begin{equation}\label{eq: estimgrad}
 \widehat{\frac{\partial U}{\partial\lambda^\learner_{\vc{x}}} }  \equiv T \sum_{n=0}^{n^{\prime}}\widehat{\Delta^\learner_{\vc{x}}(\vc{\lambda}^\learner,n) }\mathrm{Pr}[n^\learner_{\vc{x}} = n].
\end{equation}
  where estimate  $\widehat{\Delta^\learner_{\vc{x}}(\vc{\lambda}^\learner,n)}$ is constructed via sampling as follows. At each iteration, we generate $\nsamp$ samples $\vc{n}^{\learner,j}$, $j = 1,\dots, \nsamp$ of the random vector $\vc{n}^\learner$ according to the  Poisson distribution in Eq.~\eqref{eq:prod}, parameterized by the current solution vector $\vc{\lambda}^\learner$. We then compute the  empirical average:
\begin{equation}\label{eq: sampling delta}
    \widehat{\Delta^\learner_{\vc{x}}(\vc{\lambda}^\learner,n)} = \frac{1}{\nsamp}\sum_{j=1}^{\nsamp}\big(G^\learner\left(\vc{n}^{\learner,j}|_{n_{\vc{x}}^{\learner,j} = n+1}\big) \!-\! G^\learner\big(\vc{n}^{\learner,j}|_{n_{\vc{x}}^{\learner,j} = n}\big)\right),
\end{equation}
where $\vc{n}^{\learner,j}|_{n^{\learner,j}_{\vc{x}} = n}$ is equal to vector $\vc{n}^{\learner,j}$ with  $n^{\learner,j}_{\x}$ set to $n$. 

\smallskip
\noindent\textbf{Theoretical Guarantee}. Extending the analysis of  \cite{bian2017guaranteed}, and using Theorem~\ref{Thm: dr-submodular}, 
we show that the Frank-Wolfe variant combined with  gradients estimated ``well enough'' yields  a solution within a constant approximation factor from the optimal:

\begin{thm}\label{thm: final result}
Let 
\begin{align}\lambda_\mathrm{MAX} &\equiv \max_{\vc{\lambda}\in\mathcal{D}}\sum_{\learner\in\learners}||\vc{\lambda}^{\learner}||_1, \text{ and}\label{eq: lmax}\\
G_\mathrm{MAX}& \equiv \max_{\learner\in\learners, \vc{x}\in\xset}(G^\learner(\vc{e}_{\vc{x}})-G^\learner(\vc{0}))\label{eq: gmax},
\end{align}
where $e_{\vc{x}}$ is the canonical basis. Then, for any  $0<\epsilon_0,\epsilon_1<1$ and $\epsilon_2>0$, there exists a $\delta>0$ such that Algorithm \ref{alg: F-W}  terminates in at most $$K=O\big(( \frac{\sqrt{2}}{2}|\xset||\learners|T \lambda_\mathrm{MAX}^2  +2\mathcal{\lambda}_\mathrm{MAX})G_\mathrm{MAX}/\epsilon_2\big)$$ iterations, and uses  $n' = O( \lambda_\mathrm{MAX} T + \ln \frac{1}{\epsilon_1})$ terms and  $\nsamp = O(T^2n'K^2 \ln \frac{|\xset||\learners|K}{\epsilon_0})$ samples in estimator \eqref{eq: estimgrad}, so that with probability
 $1-\epsilon_0$, 
the output solution $\vc{\lambda}^K\in \mathcal{D}$ satisfies:
\begin{equation}\label{eq: final guarantee}
    U(\vc{\lambda}^K) \ge (1 - e^{\epsilon_1-1}) \max_{\vc{\lambda}\in\mathcal{D}}U(\vc{\lambda}) - \epsilon_2.
\end{equation}
\end{thm}
\yy{The proof can be found in Section \ref{proof of guarantee}.} Theorem~\ref{thm: final result} implies that, through an appropriate (but polynomial) selection of the total number of iterations $K$, the number of terms $n'$ and samples $\nsamp$, we can obtain a solution $\vc{\lambda}$ that is within $1-e^{-1}\approx 0.63$ from the optimal. The proof crucially relies on (and exploits) the continuous DR-submodularity of objective $U$, in combination with an analysis of the quality of our gradient estimator, given by Eq.~\eqref{eq: estimgrad}. 

\subsection{Proof of Theorem \ref{Thm: dr-submodular}}
\label{proof dr-submodular}
By the law of total expectation, we have:
$$
    U^\learner(\vc{\lambda}^\learner) 
     = \sum_{n=0}^\infty \mathbb{E}\left[G^\learner(\vc{n}^\learner)|n^\learner_{\vc{x}}= n\right]\cdot \frac{(\lambda^\learner_{\vc{x}} T)^{t} e^{-\lambda^\learner_{\vc{x}} T}}{t!}.    
$$
Notably, 
$\frac{\partial U}{\partial\lambda^\learner_{\vc{x}}} = \frac{\partial U^\learner(\vc{\lambda}^\learner)}{\partial \lambda^\learner_{\vc{x}}}$, for which 
the following is true: 
\begin{align*}
    \frac{\partial U^\learner(\vc{\lambda}^\learner)}{\partial \lambda^\learner_{\vc{x}}} & = \sum_{n=0}^\infty \mathbb{E}\left[G^\learner(\vc{n}^\learner)|n^\learner_{\vc{x}}= n\right]\cdot (\frac{n}{\lambda^\learner_{\vc{x}}}-T)\frac{(\lambda^\learner_{\vc{x}} T)^{n} e^{-\lambda^\learner_{\vc{x}} T}}{n!}\notag \\
    &  = \sum_{n=0}^\infty\Delta^\learner_{\vc{x}}(\vc{\lambda}^\learner,n)\cdot T\cdot \mathrm{Pr}[n^\learner_{\vc{x}} = n]
     \geq 0, 
\end{align*}
where the last inequality is true because $G$ is monotone-increasing (Lemma \ref{lem: supermodularity}). 

Next, we compute the second partial derivatives $\frac{\partial^2 U}{\partial\lambda^\learner_{\vc{x}}\partial\lambda^{\learner^\prime}_{\vc{x}^\prime}}$. 
It is easy to see that for $\textstyle \learner\neq\learner^\prime$, we have $$\frac{\partial^2 U}{\partial\lambda^\learner_{\vc{x}}\partial\lambda^{\learner^\prime}_{\vc{x}^\prime}} = 0.$$ \\
For $\learner=\learner^\prime$ and $\vc{x}=\vc{x}^\prime$, it holds that $\frac{\partial^2 U}{\partial (\lambda^\learner_{\vc{x}})^2} = \frac{\partial ^2 U^\learner(\vc{\lambda}^\learner)}{\partial (\lambda^\learner_{\vc{x}})^2}$, where
\begin{align*}
    &\frac{\partial ^2 U^\learner(\vc{\lambda}^\learner)}{\partial (\lambda^\learner_{\vc{x}})^2} =
     \Delta^\learner_{\vc{x}}(\vc{\lambda}^\learner,0)\cdot T^2e^{-\lambda^\learner_{\vc{x}} T}  \notag
    + \sum_{n=1}^\infty\Delta^\learner_{\vc{x}}(\vc{\lambda}^\learner,n)\\
    & \textstyle \left(\frac{(\lambda^\learner_{\vc{x}})^{n-1}T^{n+1}}{(n-1)!} - \frac{(\lambda^\learner_{\vc{x}})^nT^{n+2}}{n!}\right) e^{-\lambda^\learner_{\vc{x}} T}\notag\\
    =&\sum_{n=0}^\infty ( \Delta^\learner_{\vc{x}}(\vc{\lambda}^\learner,n+1) -\Delta^\learner_{\vc{x}}(\vc{\lambda}^\learner,n))\cdot\mathrm{Pr}[n^\learner_{\x} = n]T^2 
    \leq 0,
\end{align*}
and the last equality follows from the DR-submodularity of $G$ (Lemma \ref{lem: supermodularity}).

For $\learner=\learner^\prime$ and $\vc{x}\neq\vc{x}^\prime$, it holds that $\frac{\partial^2 U}{\partial \lambda^\learner_{\vc{x}}\partial \lambda^\learner_{\vc{x}^{\prime}}} = \frac{\partial ^2 U^\learner(\vc{\lambda}^\learner)}{\partial \lambda^\learner_{\vc{x}}\partial \lambda^\learner_{\vc{x}^\prime}}$, 
\begin{align}
\label{eq: hessian_notequal}
    & \frac{\partial ^2 U^\learner(\vc{\lambda}^\learner)}{\partial \lambda^\learner_{\vc{x}}\partial \lambda^\learner_{\vc{x}^\prime}}
	= \sum_{n=0}^\infty\sum_{k=0}^\infty \left(\left(\mathbb{E}\left[G^\learner(\vc{n}^\learner)|n^\learner_{\vc{x}}= n+1, n^\learner_{\vc{x}^\prime} = k+1\right]\right. \right.\notag\displaybreak[0]\\
    & - \left. \mathbb{E}\left[G^\learner(\vc{n}^\learner)|n^\learner_{\vc{x}}= n, n^\learner_{\vc{x}^\prime} = k+1\right]\right)  \notag
    - \left(\mathbb{E}[G^\learner(\vc{n}^\learner)| \right.\displaybreak[0]\\
    & n^\learner_{\vc{x}}= n+1, n^\learner_{\vc{x}^\prime} = k]  - \left. \left.\mathbb{E}\left[G^\learner(\vc{n}^\learner)|n^\learner_{\vc{x}}= n, n^\learner_{\vc{x}^\prime} = k\right]\right) \right) \notag\displaybreak[0]\\
    &\cdot \mathrm{Pr}[n^\learner_{\x} = n]\mathrm{Pr}[n^\learner_{\x'} = k]T^2
    \leq 0,
\end{align}
where the last inequality follows from the DR-submodularity of $G$ (Lemma \ref{lem: supermodularity}). \hspace{\stretch{1}} \qed

\subsection{Proof of Theorem~\ref{thm: final result}}
\yy{\label{proof of guarantee}
Our proof relies on a series of key lemmas; \arxiv{we state them below.}{we state them below, along with proof sketches.} Full  proofs of all lemmas can be found in \arxiv{the appendix}{the extended version of our paper \cite{liu2022Experimental}}.}
We begin by associating the approximation guarantee of Algorithm~\ref{alg: F-W} to the quality of gradient estimation $\widehat{\nabla U(\cdot)}$:

\begin{lem}\label{thm: final result brief}
Suppose we can construct an estimator $\widehat{\nabla U(\vc{\lambda}^k)}$ of the gradient $\nabla U(\vc{\lambda}^k)$ at each iteration $k$ such that
\begin{equation}\label{eq: estimator guarantee short}
    \langle \vc{v}^k,\nabla U(\vc{\lambda}^k)\rangle 
    \geq a \cdot\max_{\vc{v}\in\mathcal{D}}\langle \vc{v},\nabla U(\vc{\lambda}^k)\rangle - b,
\end{equation}
where 
$\vc{v}^k$ is the update direction determined by \eqref{eq: vk update}, 
$a\in(0,1]$ and $b$ are positive constants. 
Then, the output solution $\vc{\lambda}^K$ of Algorithm \ref{alg: F-W} satisfies $\vc{\lambda}^K\in\feasibleset,$ and
\begin{equation}\label{eq: final result short}
    ~U(\vc{\lambda}^K) \ge (1 - e^{-a} ) \max_{\vc{\lambda}\in\mathcal{D}}U(\vc{\lambda}) -\frac{L}{2}\lambda_{\mathrm{MAX}}\delta - b,
\end{equation}
where $L = \sqrt{2}p|\mathcal{L}|T G_\mathrm{MAX}$ 
is the Lipschitz  constant of $\nabla U$, and $\lambda_\mathrm{MAX}$ and $G_\mathrm{MAX}$ given by \eqref{eq: lmax} and \eqref{eq: gmax}. 
\end{lem}
\arxiv{The proof, found in Appendix~\ref{proof: result brief}, relies on the continuous DR-submodularity of $U$, and follows \cite{bian2017guaranteed}; we deviate from their proof to handle the additional issue that $\feasibleset$ is not downward closed 
(an assumption  in \cite{bian2017guaranteed}).
 }{\noindent\yy{\textbf{Proof Sketch.} We rely on the non-decreasing continuous DR-submodularity of $U$ (by Theorem \ref{Thm: dr-submodular}),  following the proof of Lemma 1 in \cite{bian2017guaranteed}; we deviate from their proof to handle the additional issue that $\feasibleset$ is not downward closed 
(an assumption  in \cite{bian2017guaranteed}) mainly by exploiting the fact that:
\begin{align*}
    a \max_{\vc{v}\in\mathcal{D}}\langle \vc{v},\nabla U(\vc{\lambda})\rangle - b \geq & a \langle \vc{\lambda}^*,\nabla U(\vc{\lambda})\rangle - b \\
    \geq & a\langle \vc{v}^*,\nabla U(\vc{\lambda})\rangle - b,
\end{align*}
where $\vc{\lambda}^*\in\mathcal{D}$ is the optimal solution, and $\vc{v}^*$ is a carefully constructed point, where $\vc{0}\leq \vc{v}^*\leq \vc{\lambda}^*$. 
$\hfill\qed$}}

Next, we turn our attention to characterizing the quality of our gradient estimator. To that end, use the following subexponential tail bound:  
\begin{lem}[Theorem 1 in \cite{ccanonne2017note}]\label{lem: poisson tail}
Let $n^\learner_{\vc{x}}\sim\mathrm{Poisson}(\lambda^\learner_{\vc{x}}T)$, for $\lambda^\learner_{\vc{x}}, T> 0$. Then, for any $z > \lambda^\learner_{\vc{x}}T$, we have
\begin{equation}
\label{eq: poission tail}
    \mathrm{Pr}[n^\learner_{\vc{x}}\geq z]\leq e^{-\frac{(z - \lambda^\learner_{\vc{x}}T)^2}{2\lambda^\learner_{\vc{x}}T}h(\frac{z - \lambda^\learner_{\vc{x}}T}{\lambda^\learner_{\vc{x}}T})},
\end{equation}
where $h: [-1,\infty)\to \mathbb{R}$ is the function defined by $h(u) = 2\frac{(1+u)\ln{(1+u)}-u}{u^2}$.
\end{lem}
The expression for $h(u)$ implies that the Poisson tail decays slightly faster than a standard exponential random variable (by a logarithmic factor). This lemma allows us to characterize the effect of truncating Eq.~\eqref{eq: partial derivative} in  estimation quality. In particular, 
for $n'\geq \lambda^\learner_{\x} T$,  let:
\begin{equation}\label{eq: head}
    \mathrm{HEAD}^\learner_{\vc{x}}(n') \equiv T \sum_{n=0}^{n^{\prime}}\Delta^\learner_{\vc{x}}(\vc{\lambda}^\learner,t) \mathrm{Pr}[n^\learner_{\vc{x}} = n].
\end{equation}
Then,  this is guaranteed to be within a constant factor from the true partial derivative:
\begin{lem}\label{lem: HEAD bound}
For $h(u) = 2\frac{(1+u)\ln{(1+u)}-u}{u^2}$ and $n'\geq \lambda^\learner_{\vc{x}} T$, we have:
\begin{align}
    \mathrm{HEAD}^\learner_{\vc{x}}(n')
    &\geq 
    (1 - e^{-\frac{(n^{\prime} - \lambda^\learner_{\vc{x}}T + 1)^2}{2\lambda^\learner_{\vc{x}}T}h(\frac{n^{\prime} - \lambda^\learner_{\vc{x}}T + 1}{\lambda^\learner_{\vc{x}}T})})\frac{\partial U}{\partial\lambda^\learner_{\vc{x}}}.
\end{align}
\end{lem}
\arxiv{\noindent The proof can be found in Appendix \ref{proof: HEAD bound}.}{\noindent\yy{\textbf{Proof Sketch.} The lemma follows directly from the submodularity of $G$ (Lemma~\ref{lem: supermodularity}), along with the Poisson tail bound  (Lemma~\ref{lem: poisson tail}). \hspace{\stretch{1}}$\qed$ }}
Next, by estimating $\Delta^\learner_{\vc{x}}(\vc{\lambda}^\learner,n)$ via sampling (see \eqref{eq: sampling delta}), we construct our final estimator given by \eqref{eq: estimgrad}. Putting together Lemma~\ref{lem: HEAD bound} and along with a Chernoff bound \cite{alon2004probabilistic}, to attain a guarantee on sampling, we can bound the  quality of our gradient estimator: 
\begin{lem}\label{lem: estimation guarantee} At each iteration $k$,  with probability greater than $1 - 2p|\mathcal{L}|\cdot e^{-\delta^2N/2T^2(n^{\prime}+1)}$,
\begin{equation}\label{eq: estimator guarantee full}
    \langle \vc{v}^k,\nabla U(\mathcal{\vc{\lambda}}^k)\rangle \geq a\cdot\max_{\vc{v}\in\mathcal{D}}\langle \vc{v},\nabla U(\mathcal{\vc{\lambda}}^k)\rangle - b,
\end{equation}
where 
\begin{align}
    &a = 1- \max_{k=1,\dots, K}\mathrm{P}^k_\mathrm{MAX}, \quad \text{and}\\
    &b = 2\lambda_{\mathrm{MAX}}\delta\cdot G_\mathrm{MAX},
\end{align}
for $\mathrm{P}^k_\mathrm{MAX} = \max_{l\in\learners,\x\in\xset} \mathrm{P}[n^{\learner,k}_{\x}\geq n^{\prime} + 1]$ ($n^{\learner,k}_{\x}$ is a Poisson r.v. with parameter $\lambda^{\learner,k}_{\x}T$), and with $\lambda_\mathrm{MAX}$ and $G_\mathrm{MAX}$ given by Eq. \eqref{eq: lmax} and \eqref{eq: gmax}. 
\end{lem}
\arxiv{\noindent The proof is in Appendix~\ref{proof: estimation guarantee}.}{%
\noindent\yy{\textbf{Proof Sketch.} We again follow the proof of Lemma 3.2 in \cite{calinescu2011maximizing}. Utilizing Chernoff bounds described by Theorem A.1.16 in \cite{alon2004probabilistic}, and the constructed auxiliary variables, we calculate bounds for the distance between $\mathrm{HEAD}$ and final estimator of the partial derivative. By Lemma~\ref{lem: HEAD bound}, we further calculate bounds for the distance between the true partial derivative and final estimator of the partial derivative, which in turn imply~\eqref{eq: estimator guarantee full}.  $\hfill\qed$ 
}}

\yy{Theorem~\ref{thm: final result} follows by combining Lemmas~\ref{thm: final result brief} and \ref{lem: estimation guarantee}. In particular,}
by Lemma \ref{lem: estimation guarantee} and a union bound, we have that \eqref{eq: estimator guarantee full} is satisfied for all iterations with probability greater than $1 - 2|\xset||\mathcal{L}|\cdot e^{-\delta^2N/2T^2(n^{\prime}+1)}$. This, combined with Lemma~\ref{thm: final result brief}, implies that
\begin{align*}
    U(\vc{\lambda}^K)\ge & (1 - e^{\mathrm{P_{MAX}}-1}) \cdot\max_{\vc{\lambda}\in\mathcal{D}}U(\vc{\lambda}) \\
    &\textstyle -( \frac{\sqrt{2}}{2}|\xset||\learners|T \lambda_\mathrm{MAX}^2
     + 2\mathcal{\lambda}_\mathrm{MAX})G_\mathrm{MAX}\delta,
\end{align*}
is satisfied with the same probability.
This implies that for any $0<\epsilon_0,\epsilon_1<1$ and $\epsilon_2>0$, $$U(\vc{\lambda}^K) \ge (1 - e^{\epsilon_1-1}) \cdot\mathrm{OPT} - \epsilon_2,$$
with probability $1-\epsilon_0$.
From Eq. \eqref{eq: poission tail}, the probability is an increasing function w.r.t. $\lambda^\learner_{\vc{x}}$, and $\lambda_\mathrm{MAX}$ is an upper bound for $\lambda^\learner_{\vc{x}}$. Letting  $$u  = \frac{n' - \lambda_\mathrm{MAX}T}{\lambda_\mathrm{MAX}T},$$ we have 
\begin{align*}
\mathrm{P_{MAX}} & \le e^{-\frac{(n' - \lambda_\mathrm{MAX}T)^2}{2\lambda_\mathrm{MAX}T}h(\frac{n' - \lambda_\mathrm{MAX}T}{\lambda_\mathrm{MAX}T})} \\
& = e^{-\lambda_\mathrm{MAX}T( (1+u)\ln(1+u) -u )}
= \Omega(e^{-\lambda_\mathrm{MAX}T u}) = \epsilon_1,
\end{align*}
where the last line holds because $u \ln u - u > u$ when $u$ is large enough, e.g., $u\ge e^2$. Thus, $n' = O( \lambda_\mathrm{MAX} T + \ln \frac{1}{\epsilon_1})$. \\
We determine $K$ and $N$  by setting $$( \frac{\sqrt{2}}{2}|\xset||\learners|T \lambda_\mathrm{MAX}^2  +2\mathcal{\lambda}_\mathrm{MAX})G_\mathrm{MAX}/K = \epsilon_2$$ and $$2|\xset||\mathcal{\learners}|K\cdot e^{-N/2T^2(n^{\prime}+1)K^2} = \epsilon_0.$$
Therefore, $K=O(( \frac{\sqrt{2}}{2}|\xset||\learners|T \lambda_\mathrm{MAX}^2  +2\mathcal{\lambda}_\mathrm{MAX})G_\mathrm{MAX}/\epsilon_2)$, and $N = O(T^2n'K^2 \ln \frac{|\xset||\learners|K}{\epsilon_0})$. 
$\hfill\qed$

\section{Extensions}\label{sec: extension}
Our model extends  in many ways (e.g., to multiple types per learner). 
We discuss three non-trivial extensions below. 

\noindent\textbf{Heterogeneous Noisy Sources.}
Our model and analysis directly generalizes to a heterogeneous (or \emph{heteroskedastic}) noise setting, in which the noise level varies across sources. Formally, labels of type $\type$ at source $\source$ are generated via $y = \vc{x}^\top\vc{\beta}_\type + \epsilon_{\type,\source}$, where $\epsilon_{\type,\source}$ are zero-mean normal noise variables with variance $\sigma^2_{\type,\source}$. In this case, the estimator in \eqref{eq:map} needs to be replaced by Generalized Least Squares \cite{friedman2001elements}, whereby every pair $(\x,y)\in\reals^d\times \reals$  of type $\type$ generated by  $\source$ is replaced by $(\frac{\x}{\sigma_{\type,\source}},\frac{y}{\sigma_{\type,\source}})\in\reals^d\times \reals$ prior to applying Eq.~\eqref{eq:map}. This, in turn, changes the D-optimality criterion objective, so that $\sigma_\type^2$ is replaced by $\sigma_{\type,\source}^2$ for vectors $\vc{x}\in \xset$ coming from source $\source$. In other words, data coming from noisy sources are valued less by the learner. This rescaling preserves the monotonicity and continuous DR-submodularity of our overall objective, and our guarantees  hold, \emph{mutatis mutandis}.

\noindent\textbf{Uncountable $\xset$.} We assumed in our analysis that data features are generated from a finite set $\xset$, and that transmission rates per edge are parametrized by both the type $\type\in\types$ \emph{and} the features $\x$ of the data pair transmitted. This a priori prevents us from considering an infinite set of experiments $\xset$: this would lead to an infinite set of constraints in Problem~\eqref{prob: utility}. \si{In practice, it would also make routing  intractable, as routing decisions depend on both $\type$ and $\x$.}

We can however extend our analysis to a setting where experiments $\xset$ are infinite, or even uncountable. To do so, we can consider rates  per edge $e$ of the form $\lambda_{\source,\type}^e$, i.e., parameterized by type $\type$ and \emph{source} $s$ rather than features $\x$. In practice, this would mean that packets would be routed based on the source and type, not inspecting the features of the internal pairs, while constraints would be finite (depending on $|\sources|$, not $|\xset|$). Data generation at source $\source$ can then be modelled via a compound Poisson process with rate $\lambda_{\source,\type}$, at the epochs of which the features $\x$ are sampled independently from some probability distribution $\nu_{\source,\type}$ over $\reals^d$. The objective then would be written as an expectation over not only arrivals at a learner from source $\source$ (which will again be Poisson) but also the distribution $\nu_{\source,\type^{\learner}}$ of features. Sampling from the latter would need to be used when estimating $\nabla U$; as long as  Chernoff-type bounds can be used to characterize the estimation quality of such sampling (which would be the case if, e.g., $\nu_{\source,\type}$ are Gaussian), our analysis would still hold, taking also the number of sampled features into account.



\noindent\textbf{Arbitrary (Non-DAG) Topology.}
For notational convenience, we assumed that graph $G$ was a DAG, with sources and sinks corresponding to sets $\sources$ and $\learners$ respectively. 
Our analysis \si{further} 
extends to more general \si{(i.e., non-DAG)} graphs, 
provided that extra care is taken for flow constraints to prevent cycles. \si{This can be accomplished, e.g., via source routing. Given an arbitrary graph, and arbitrary locations for data sources and learners, we can extend our setting as follows: }(a) flows from a source $\source$ to a learner $\learner$ could follow source-path routing, over one or more directed paths linking the two, and (b) flows could be indexed by (and remain constant along) a path, in addition to $\x$ and $\type$, while also ensuring that  (c) aggregate flow across all paths that pass through an edge does not violate capacity constraints. Such a formulation still yields a linear set of constraints, and our analysis still holds. In fact, in this case, the corresponding set $\feasibleset$ is downward closed, so  the proof of the corresponding Theorem~\ref{thm: final result brief} follows more directly from \cite{bian2017guaranteed}. 

\section{Numerical Evaluation}\label{sec: simulation}

\begin{table}[t]
\caption{Graph Topologies and Experiment Parameters}
\centering
\begin{tabular}{cccccccc}
Graph & $|V|$ & $|E|$ & $|\xset|$ & $|\types|$ & $|\sources|$ & $|\learners|$ & $U_{\texttt{FW}}$\\
\hline
\multicolumn{8}{c}{synthetic topologies}\\
\hline
Erd\H{o}s-R\'enyi (\texttt{ER}) & 100 & 1042 & 20 & 5 & 10 & 5 & 309.95\\
balanced-tree (\texttt{BT}) & 341 & 680 & 20 & 5 & 10 & 5 & 196.68 \\
hypercube (\texttt{HC}) & 128 & 896 & 20 & 5 & 10 & 5 & 297.69 \\
\texttt{star} & 100 & 198 & 20 & 5 & 10 & 5 & 211.69 \\
\texttt{grid} & 100 & 360 & 20 & 5 & 10 & 5  & 260.12\\
small-world (\texttt{SW})~\cite{kleinberg1999small} & 100 & 491 & 20 & 5 & 10 & 5 & 272.76 \\
\hline
\multicolumn{8}{c}{real backbone networks~\cite{rossi2011caching}} \\
\hline
\texttt{GEANT} & 22 & 66 & 20 & 3&  3 & 3 & 214.30 \\ 
\texttt{Abilene} & 9 & 26 & 20 & 3&  3 & 3 & 216.88 \\
Deutsche Telekom  & \multirow{2}{*}{68} & \multirow{2}{*}{546} & \multirow{2}{*}{20} & \multirow{2}{*}{3} & \multirow{2}{*}{3} & \multirow{2}{*}{3} & \multirow{2}{*}{232.52} \\
(\texttt{Dtelekom}) \\
\end{tabular}
\label{topologies}
\end{table}

To evaluate the proposed algorithm, we perform simulations over a number of network topologies and with several different network parameter settings,
summarized in Table \ref{topologies}.

\subsection{Experiment Setting}
We consider a finite feature set $\xset$ that includes randomly generated feature vectors with $d=100$, and a set $\types$ that of different Bayesian linear regression models with
$\vc{\beta}_\type$, $\type\in\types$. Labels of each type are generated with Gaussian noise, whose variance $\sigma_{\type}$ is uniformly at random (u.a.r.) chosen from 0.5 to 1.
For each network, we u.a.r. select $|\learners|$ learners 
\blu{and $|\sources|$ data sources, and remove incoming edges of  sources and outgoing edges of  learners. Each learner has a target model $\vc{\beta}_{\type^\learner}$, $\type^\learner\in\types$ with a diagonal prior $\vc{\Sigma}_{\type^\learner}$ generated as follows. First, we separate features into two classes: well-known and poorly-known. Then, we set the  corresponding prior covariance (i.e., the diagonal elements in  $\vc{\Sigma}_{\type^\learner}$) to low (uniformly from 0 to 0.01) and high (uniformly from 100 to 200) values,  for well-known and poorly-known features, respectively.} 
The link capacity $\mu^e, e=(u,v)\in\mathcal{E}$ is selected u.a.r. from 50 to 100, and source $s$ generates the data $(\x, y)$ of type $\type$ label with rate $\lambda_{\x,\type}^{\source}$, uniformly distributed over [2,5].

\noindent\textbf{Algorithms.} We compare our proposed Frank-Wolfe based algorithm (we denote it by \texttt{FW}) with several baseline data transmission strategies derived in different ways:
\begin{itemize}
	\item \texttt{MaxSum:} This maximizes the aggregate total useful incoming traffic rates of learners, i.e., the objective is: $$U_{\texttt{MaxSum}}(\vc{\lambda})=\sum_{\learner\in\learners}\sum_{\x\in\xset}\lambda^\learner_{\x}.$$ 
	\item \texttt{MaxAlpha:} This maximizes the aggregate $\alpha$-fair utilities \cite{srikant2012mathematics} of the total useful incoming traffic at learners, i.e., the objective is: $$U_{\texttt{MaxAlpha}}(\vc{\lambda})=\sum_{\learner\in\learners}(\sum_{\x\in\xset}\lambda^\learner_{\x})^{1-\alpha}/(1-\alpha).$$ We set $\alpha = 5.$
\end{itemize}
We also compare with another algorithm for the proposed experimental design networks:
\begin{itemize}
	\item \texttt{PGA:} it also solves Prob.~\eqref{prob: utility}, as does our proposed algorithm, via the projected gradient ascent~\cite{hassani2017gradient}. As \texttt{PGA} also requires gradients, we use our novel gradient estimation (by Eq. \eqref{eq: estimgrad}).
\end{itemize}

Note that projected gradient ascent finds a solution within $1/2$ from the optimal if the true gradients are accessible~\cite{hassani2017gradient}; its theoretical guarantee with estimated gradients is out of the scope of this work.

\begin{figure}[t]
\centering
\subfigure[Normalized Aggregate Utility]{
\begin{minipage}{1\linewidth}
\centering
\includegraphics[width=1.0\linewidth]{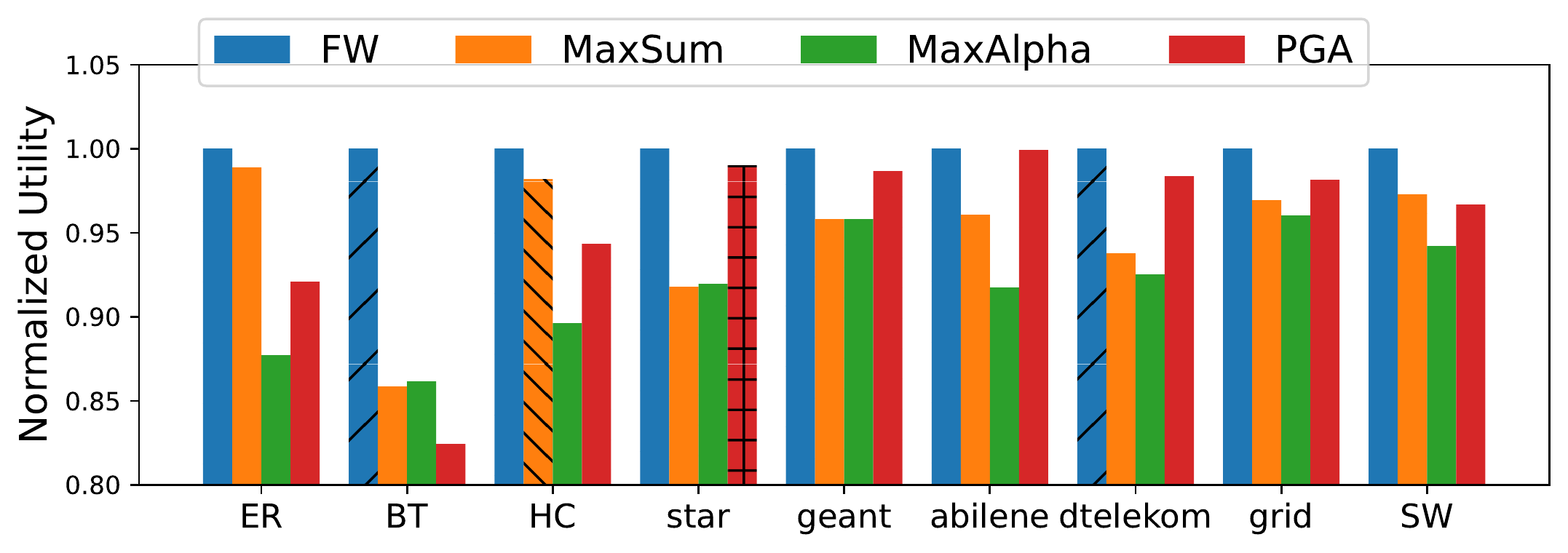}
\label{fig: topologies utility}
\end{minipage}
}
\subfigure[Average Norm of Estimation Error per Learner]{
\begin{minipage}{1\linewidth}
\centering
\includegraphics[width=1.0\linewidth]{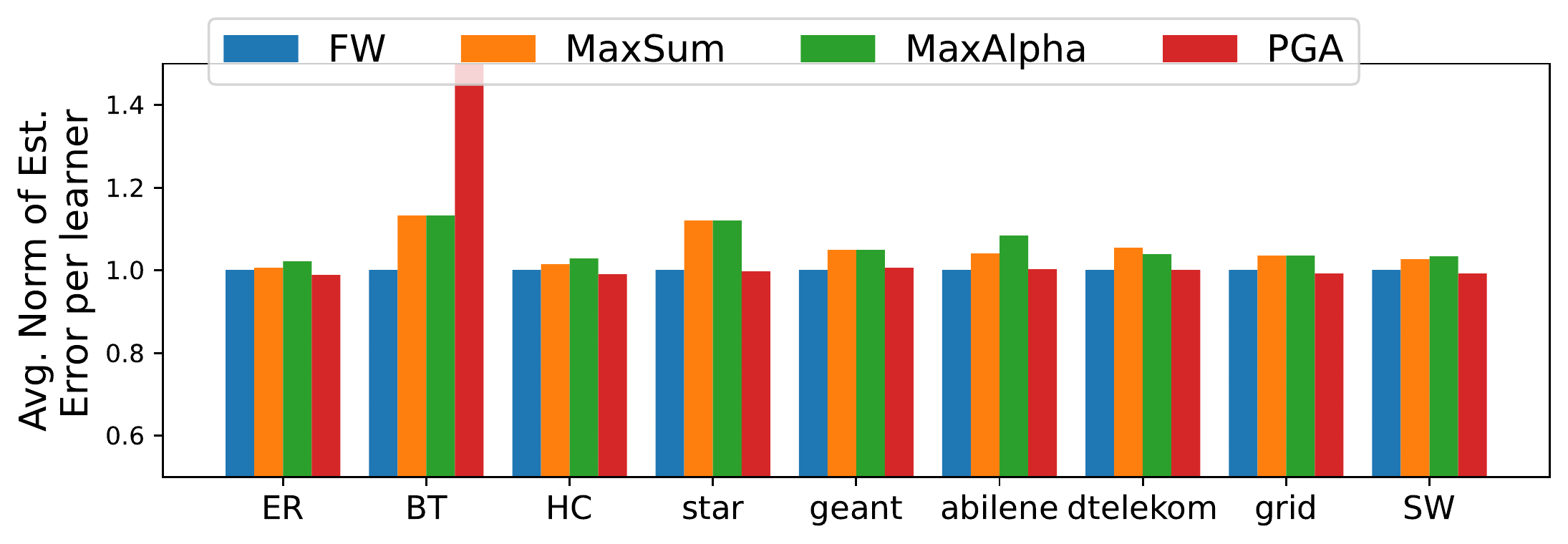}
\label{fig: topologies error}
\end{minipage}
}
\caption{Normalized aggregate utility and average norm of estimation error per leaner in different networks. Utilities are normalized by the utility of Algorithm 1 (\texttt{FW}) $U_{\texttt{FW}}$, reported in Table \ref{topologies}. We can observe that \texttt{FW} is the best in terms of maximizing the utility and minimizing the estimation error in all networks. 
}
\label{fig: different_top}
\end{figure}

\noindent\textbf{Simulation Parameters.} We run \texttt{FW} and \texttt{PGA} for $K = 100$ iterations with step size $\delta = 0.01$. In each iteration, we estimate the gradient according to Eq. \eqref{eq: estimgrad} with $N = 100$, and $n' = \lceil 2\max_{\learner,\x}\lambda^\learner_{\x}T\rceil$, where $\lambda^\learner_{\x}$ is given by the current solution. We consider a data acquisition time $T=10$. Since our objective function cannot be computed in closed-form, we rely on sampling with $1000$ samples. We also evaluate the model training quality by the average norm of estimation error, where the estimation error is the difference between the true model and the MAP estimator, given by \eqref{eq:map}. We average over 1000 realizations of the label noise as well as the number of data arrived at the learner $\{\vc{n}^\learner\}_{\learner \in \learners}$.  

\begin{figure} 
\centering 
\subfigure[Utility - Source Rates]{\label{fig: SourceU}
\includegraphics[width=0.46\linewidth]{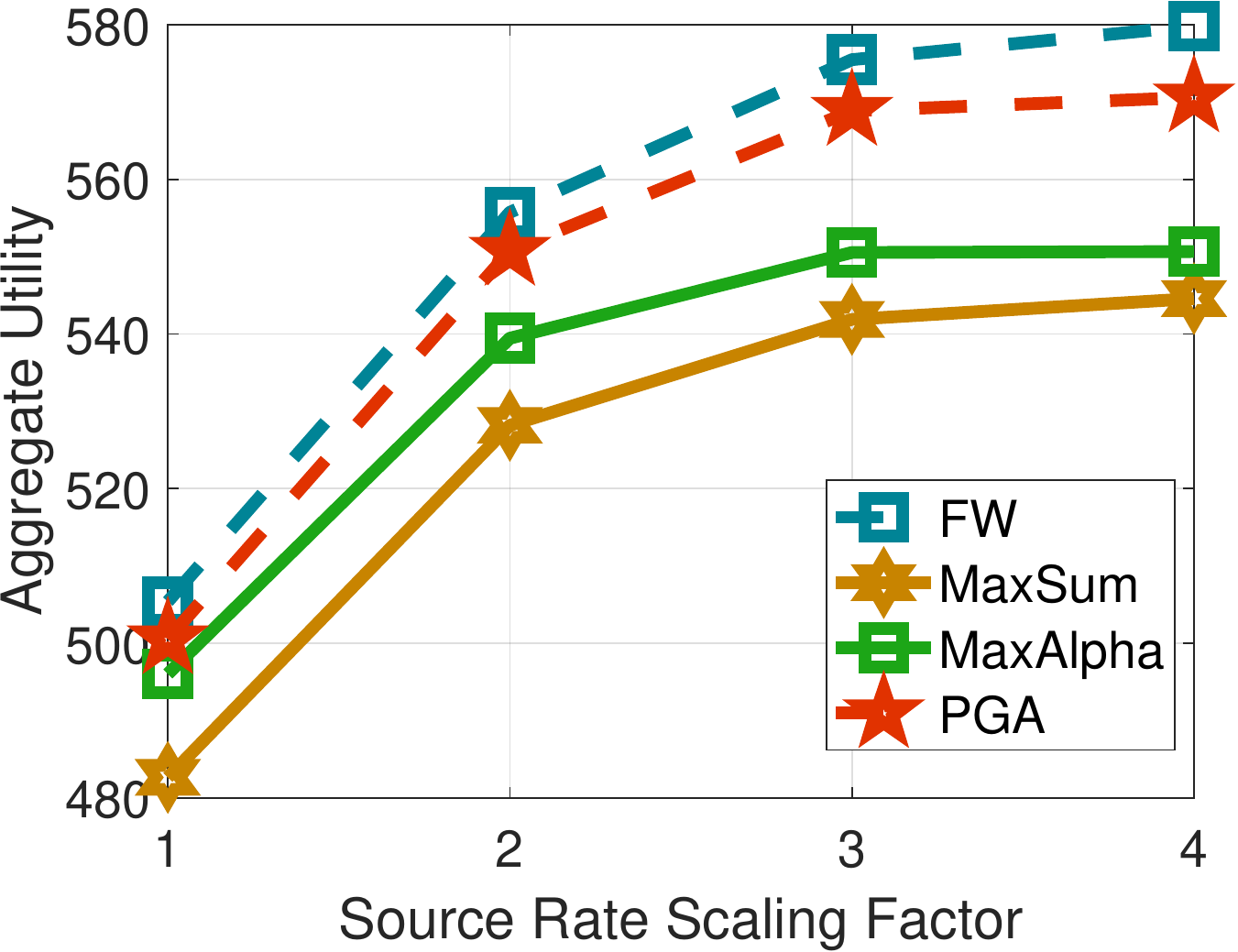}}
\hspace{0.01\linewidth}
\subfigure[Est. Error - Source Rates]{\label{fig: SourceB}
\includegraphics[width=0.46\linewidth]{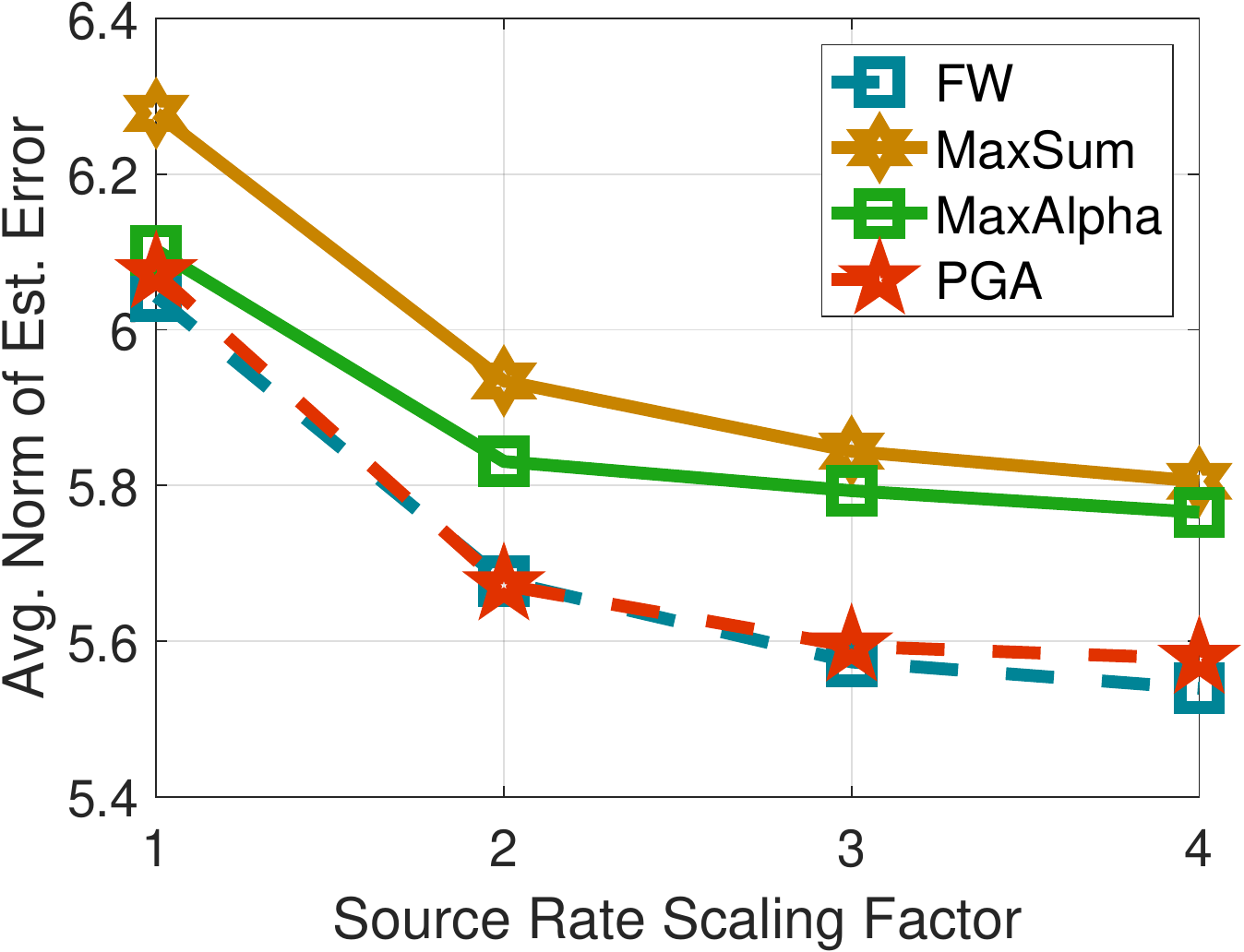}}
\vfill
\subfigure[Utility - Link Capacities]{\label{fig: CapU}
\includegraphics[width=0.46\linewidth]{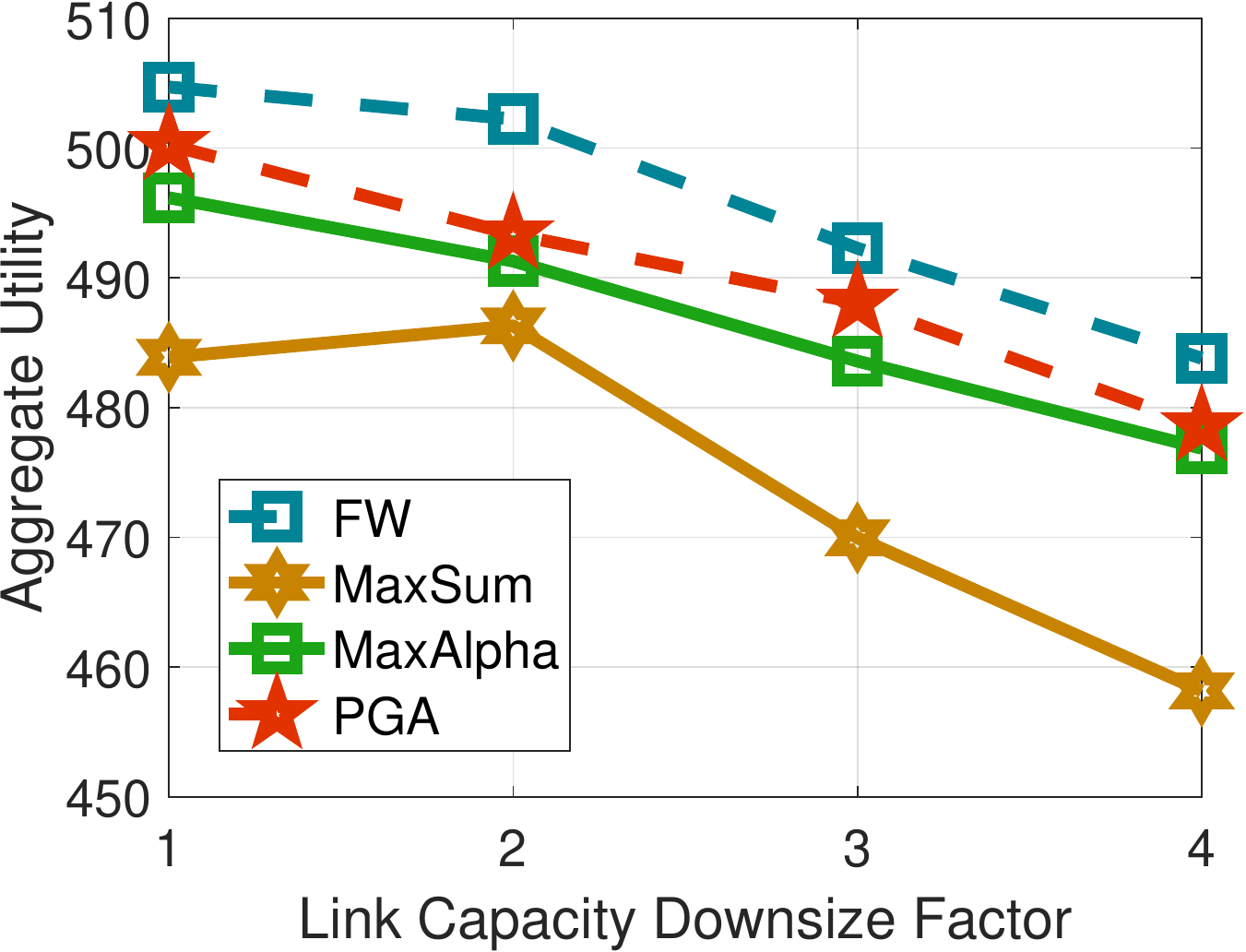}}
\hspace{0.01\linewidth}
\subfigure[Est. Error - Link Capacities]{\label{fig: CapB}
\includegraphics[width=0.46\linewidth]{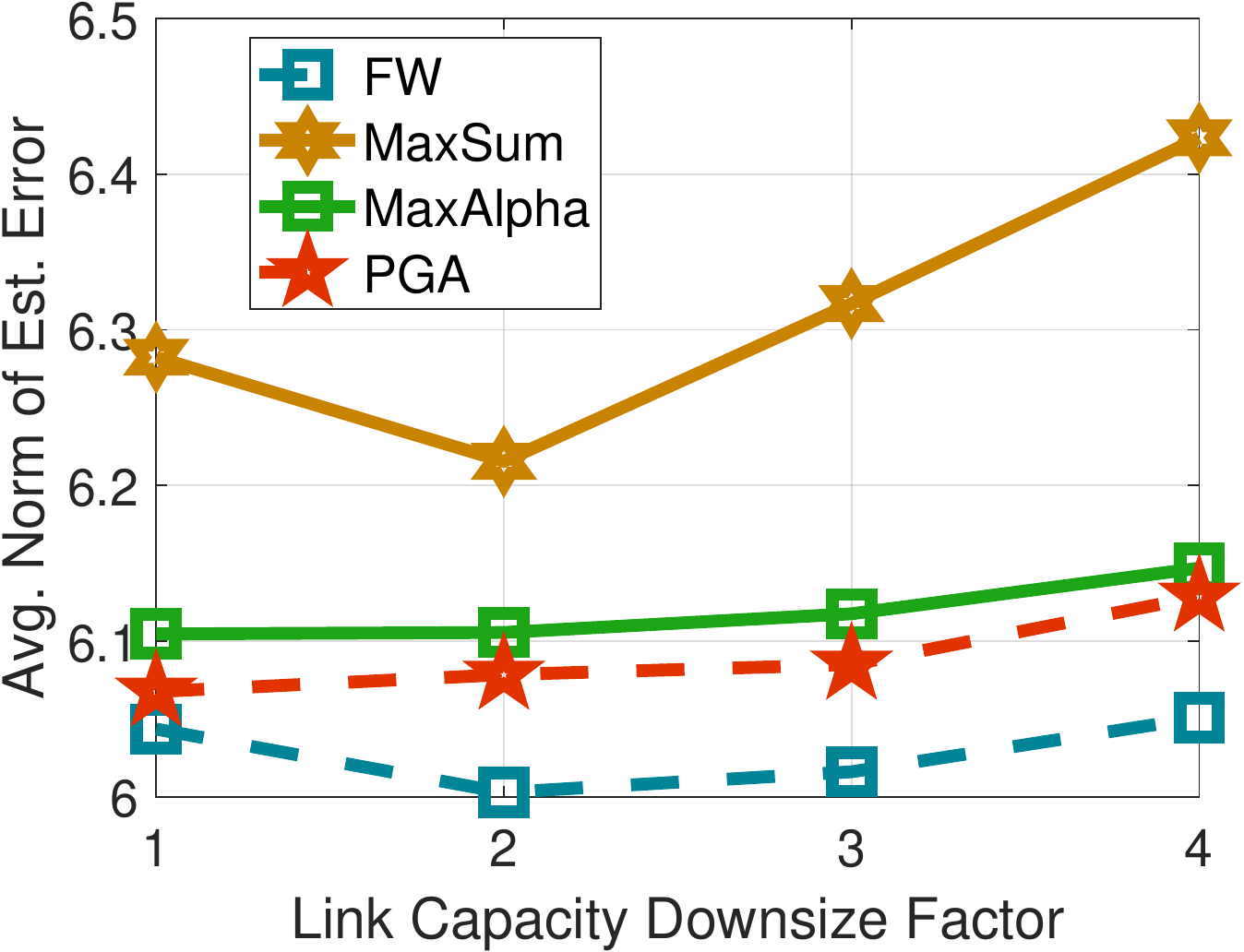}}
\caption{Algorithm comparison in abilene topology with different parameter settings. The achieved aggregate utility and model estimation quality of different algorithms are evaluated with different source data rates and bottleneck link capacities.}
\label{fig: Abilene parameters}
\end{figure}

\subsection{Results}
\noindent\textbf{Performance over Different Topologies.} We first compare the proposed algorithm (\texttt{FW}) with baselines in terms of the normalized aggregated utility and model estimation quality over several networks, shown in Figures \ref{fig: topologies utility} and \ref{fig: topologies error}, respectively. Utilities in Fig. \ref{fig: topologies utility} are normalized by the aggregate utility of \texttt{FW}, reported in  Tab. \ref{topologies} under $U_{\texttt{FW}}$.  Learners in these networks have distinct target models to train. In all network topologies, \texttt{FW} outperforms \texttt{MaxSum} and \texttt{MaxAlpha} in both aggregate utility and average norm of estimation error. \texttt{PGA}, which also is based on our experimental design framework, performs well (second best) in most networks, except balanced tree, in which it finds a bad stationary point.

\noindent\textbf{Effect of Source Rates and Link Capacities.} \blu{Next, we evaluate how algorithm performance  is affected by varying source rates as well as link capacitites. We focus on the Abilene network, having 3 sources and 3 learners, where  two of the learners have a same target training model.
 We set the data acquisition time to $T=1$, and labels are generated with Gaussian noise with variance 1. Finally, we again use diagonal prior covariances, and again split between high-variance (selected uniformly between 1 and 3) and low-variance (selected uniformly between 0 and 0.1) features. }


Figures~\ref{fig: SourceU} and \ref{fig: SourceB} plot the aggregate utility and average total norm of estimation error across learners, with different data source rates at the sources. The initial source rates are sampled u.a.r. from 2  to 5, and we
scale it by different scaling factors. 
As the source rates increases, the aggregate utility increases and the norm of estimation error decreases for all algorithms. \texttt{FW} is always the best in both figures. Moreover, the proposed experimental design framework can significantly improve the training quality: algorithms based on our proposed framework (\texttt{FW} and \texttt{PGA}) with source rates scaled by $2$ already outperform the other two algorithms (\texttt{MaxSum} and \texttt{MaxAlpha}) with source rates scaled by $4$. We see reverse results of \texttt{MaxSum} and \texttt{MaxAlpha} in these two figures compared with Figure~\ref{fig: different_top}, showing that the algorithm which considers fairness (i.e., $\alpha$-fair utilities), may perform better if we have competing learners.

Figures \ref{fig: CapU} and \ref{fig: CapB} show performance in Abilene network with different link capacities of several bottleneck links. The capacities are initially sampled u.a.r. from 50 to 100, and we divide it by  different downsize factors. 
The overall trend is that as the link capacities decrease, algorithms achieve smaller aggregate network utility and get a higher average norm of estimation error. The proposed algorithm is always the best with different bottleneck link capacities in both figures.

\section{Conclusion}\label{sec: conclusion}
We propose \emph{experimental design networks}, to determine a data transmission strategy that maximizes the quality of  trained models in a distributed learning system.\footnote{\yy{Our code and data are publicly available at \href{https://github.com/neu-spiral/Networked-Learning}{https://github.com/neu-spiral/Networked-Learning}}.} 
The underlying optimization problem can be approximated even though its objective function is non-concave. 

Beyond extensions we have already listed, our framework can be used  to explore other experimental design objectives (e.g., A-optimality and E-optimality) as well as variants that include data source generation costs. Distributed and adaptive implementations of the rate allocation schemes we proposed are also  interesting future directions.
\yy{Incorporating non-linear learning tasks (e.g., deep neural networks) is also an open avenue of exploration: though Bayesian posteriors are harder to compute in closed-form for this case, techniques such as variational inference \cite{jaakkola1997variational} can be utilized to approach this problem. }
\si{Finally, an interesting extension of our model involves a multi-stage setting, in which learners receive data in one stage, update their posteriors, and use these as new priors in the next stage. Studying the dynamics of such a system, as well as how network design impacts these dynamics, is a very interesting open problem. } 

\section*{Acknowledgment}

\yy{The authors gratefully acknowledge support from the National Science Foundation (grants 1718355, 2107062, and 2112471).}


\bibliographystyle{IEEEtran}  
\bibliography{references}

\appendix

\subsection{Proof of Lemma \ref{lem: supermodularity}}\label{proof: supermodular}

\begin{proof}
We extend the proof in Appendix A of \cite{horel2014budget} for the supmodularity of D-optimal design over a set to the integer lattice: For $\vc{n}\in\mathbb{N}^{p}$ and $k\in\mathbb{N}$, we have
\begin{align*}
    G(\vc{n} &+ k\vc{e}_i) - G(\vc{n})=\\ 
    =& \log\det(\mathbf{I}_d + \frac{k}{\sigma^2}\x_i\x_i^{\top}(\Sigma_0^{-1}+\sum_{i=1}^{p} \frac{n_i}{\sigma^2}\x_i\x_i^{\top})^{-1})\\
    =&   \log(1+\frac{k}{\sigma^2}\x_i^{\top}A(\vc{n})^{-1}\x_i),
\end{align*}
where $A(\vc{n}) = \Sigma_0^{-1}+\sum_{i=1}^{p} \frac{n_i}{\sigma^2}\x_i\x_i^{\top}$, and the last equality follows Eq. (24) in \cite{petersen2012matrix}. The monotonicity of $G$ \si{follows}  because $A(\vc{n})^{-1}$ is positive semidefinite. Finally, since the \si{matrix} inverse is decreasing over the \si{positive semi-definite order}, we have $A(\vc{n})^{-1}\succeq A(\vc{m})^{-1}$, $\forall\ \vc{n}, \vc{m}\in\mathbb{N}^{p}\ \text{and}\ \vc{n}\leq\vc{m}$, which leads to $
G(\vc{n} + k\vc{e}_i) - G(\vc{n})\geq G(\vc{m} + k\vc{e}_i) - G(\vc{m})
$.
\end{proof}

\subsection{Proof of Lemma \ref{thm: final result brief}}\label{proof: result brief}
\begin{proof}
To start with, $\vc{\lambda}^K\in\mathcal{D}$, as a convex combination of points in $\mathcal{D}$. Next, consider the point $\vc{v}^*=(\vc{\lambda}^*\lor \vc{\lambda})-\vc{\lambda} = (\vc{\lambda}^*-\vc{\lambda})\lor \vc{0} \ge \vc{0}$, in which $\vc{\lambda}$ is the solution at current iteration and $\vc{\lambda}^*\in\mathcal{D}$ is the optimal solution. Because $U$ is non-decreasing (Thm. \ref{Thm: dr-submodular}), we have
\begin{equation}
\label{eq: optimal}
    U(\vc{\lambda}+\vc{v}^*) = U(\vc{\lambda}^*\lor \vc{\lambda})\ge U(\vc{\lambda}^*).
\end{equation} 
A DR-submodular continuous function is concave along any non-negative direction, and any non-positive direction (see e.g., Prop. 4 in \cite{bian2017guaranteed}), thus $g(\xi):=U(\vc{\lambda}+\xi \vc{v}^*)$, where $g'(\xi)= \langle \vc{v}^*,\nabla U(\vc{\lambda}+\xi \vc{v}^*)\rangle$, is concave, hence, 
\begin{equation}
\label{eq: concave}
    U(\vc{\lambda}^*)-U(\vc{\lambda}) = g(1)-g(0) 
	\leq g'(0)\times 1 = \langle \vc{v}^*,\nabla U(\vc{\lambda})\rangle.
\end{equation}
Then,
\begin{align}\label{eq: algorithm update quality}
    & \langle \vc{v},\nabla U(\vc{\lambda})\rangle \overset{\eqref{eq: estimator guarantee short}}{\geq}  a \max_{\vc{v}\in\mathcal{D}}\langle \vc{v},\nabla U(\vc{\lambda})\rangle - b \geq a \langle \vc{\lambda}^*,\nabla U(\vc{\lambda})\rangle - b \notag\\
    & \geq a\langle \vc{v}^*,\nabla U(\vc{\lambda})\rangle - b 
     \overset{\eqref{eq: concave}}{\geq} a(U(\vc{\lambda}+\vc{v}^*)-U(\vc{\lambda})) - b\notag\\ 
     &\overset{\eqref{eq: optimal}}{\geq} a(U(\vc{\lambda}^*)-U(\vc{\lambda})) - b,
\end{align}
where the second inequality is because the LHS maximizes the inner product and the third inequality is because $\vc{0}\leq \vc{v}^*\leq \vc{\lambda}^*$ and $\nabla U(\vc{\lambda})$ is positive (Thm. \ref{Thm: dr-submodular}). 
From the definition of the Hessian, 
we can show that $||\nabla^2 U||_F$ is bounded by $\sqrt{2}{|\xset|}|\mathcal{L}|T G_\mathrm{MAX}$ (because 2-norm is smaller than Frobenius norm \cite{horn2012matrix}), thus $L = \sqrt{2}{|\xset|}|\mathcal{L}|T G_\mathrm{MAX}$ is the Lipschitz continuous constant of $\nabla U$ . 
Then, we have
\begin{align*}
    &U(\vc{\lambda}^{k+1}) - U(\vc{\lambda}^k) = U(\vc{\lambda}^k+\gamma_k \vc{v}^k) -U(\vc{\lambda}^k) \notag\\
    & \ge \gamma_k \langle \vc{v}^k, \nabla U(\vc{\lambda}^k) \rangle - \frac{L}{2} \gamma_k^2||\vc{v}^k||_2^2 \text{ (Lipschitz) }\notag\\
    & \overset{\eqref{eq: algorithm update quality}}{\ge} a \gamma_k (\max_{\vc{\lambda}\in\mathcal{D}}U(\vc{\lambda})-U(\vc{\lambda}^k)) - \gamma_k b - \frac{L}{2} \gamma_k^2||\vc{v}^k||_2^2
\end{align*}
After rearrangement, we have
\begin{align*}
    & U(\vc{\lambda}^{k+1}) - \max_{\vc{\lambda}\in\mathcal{D}}U(\vc{\lambda})\ge \\
    & (1-a \gamma_k)[U(\vc{\lambda}^k)-\max_{\vc{\lambda}\in\mathcal{D}}U(\vc{\lambda})] - \gamma_k b - \frac{L}{2} \gamma_k^2\lambda_\mathrm{MAX}^2,
\end{align*}
since $||\vc{v}^k||_2^2\leq\lambda_\mathrm{MAX}^2$. By telescope,
\begin{align*}
    & U(\vc{\lambda}^{K}) - \max_{\vc{\lambda}\in\mathcal{D}}U(\vc{\lambda})
    \geq \\
    & [U(\vc{\lambda}^0) - \max_{\vc{\lambda}\in\mathcal{D}}U(\vc{\lambda})]e^{-a } - b - \frac{L}{2} \sum_{k=0}^{K-1} \gamma_k^2\lambda_\mathrm{MAX}^2.
\end{align*}
Finally, as $U(\vc{\lambda}^0) = 0$ and $\gamma_k = \delta = 1/K$, we have 
\begin{align*}
    U(\vc{\lambda}^K) \ge (1 - e^{-a}) U(\vc{\lambda}^*) -\frac{L}{2}\delta\lambda_\mathrm{MAX}^2 -b.
\end{align*}
\end{proof}

\subsection{Proof of Lemma \ref{lem: HEAD bound}}\label{proof: HEAD bound}
\begin{proof}
We further define
\begin{align*}
    \mathrm{TAIL}^\learner_{\vc{x}} = \sum_{n = n^{'} + 1}^{\infty}\Delta^\learner_{\vc{x}}(\vc{\lambda}^\learner,n)\cdot 
    T\cdot \mathrm{Pr}[n^\learner_{\vc{x}} = n].
\end{align*}
We have 
$$\mathrm{HEAD}^\learner_{\vc{x}}\geq\Delta^\learner_{\vc{x}}(\vc{\lambda}^\learner,n^\prime)\cdot T\cdot\mathrm{Pr}[n^\learner_{\vc{x}}\leq n^{\prime}]$$ and $$\mathrm{TAIL}^\learner_{\vc{x}}\leq\Delta^\learner_{\vc{x}}(\vc{\lambda}^\learner,n^\prime)\cdot T\cdot\mathrm{Pr}[n^\learner_{\vc{x}}\geq n^{\prime} + 1],$$
since $\Delta^\learner_{\vc{x}}(\vc{\lambda}^\learner,t_1)\geq \Delta^\learner_{\vc{x}}(\vc{\lambda}^\learner,t_2)$ for $t_1\leq t_2$, resulting from the submodularity of $G$ (Lemma \ref{lem: supermodularity}). We note that $\frac{\partial U}{\partial \lambda^\learner_{\vc{x}}} = \mathrm{HEAD}^\learner_{\vc{x}} + \mathrm{TAIL}^\learner_{\vc{x}}$. Then we have,
\begin{align*}
& \frac{\mathrm{HEAD}^\learner_{\vc{x}}}{\mathrm{HEAD}^\learner_{\vc{x}} + \mathrm{TAIL}^\learner_{\vc{x}}} = \frac{1}{1 + \frac{\mathrm{TAIL}^\learner_{\vc{x}}}{\mathrm{HEAD}^\learner_{\vc{x}}}} \geq \\
& \frac{1}{1 + \frac{\Delta^\learner_{\vc{x}}(\vc{\lambda}^\learner,n^\prime)\cdot T\cdot\mathrm{Pr}[n^\learner_{\vc{x}}\geq n^{\prime} + 1]}{\Delta^\learner_{\vc{x}}(\vc{\lambda}^\learner,n^\prime)\cdot T\cdot(1 - \mathrm{Pr}[n^\learner_{\vc{x}}\geq n^{\prime} + 1]})} = 1 - \mathrm{Pr}[n^\learner_{\vc{x}}\geq n^{\prime} + 1], 
\end{align*}
thus, 
\begin{align*}
\mathrm{HEAD}^\learner_{\vc{x}} & \geq (1 - \mathrm{Pr}[n^\learner_{\vc{x}}\geq n^{\prime} + 1])\frac{\partial U}{\partial \lambda^\learner_{\vc{x}}} \\
& \geq (1-e^{-\frac{(n^{\prime} - \lambda^\learner_{\vc{x}}T + 1)^2}{2\lambda^\learner_{\vc{x}}T}h(\frac{n^{\prime} - \lambda^\learner_{\vc{x}}T + 1}{\lambda^\learner_{\vc{x}}T})})\frac{\partial U}{\partial \lambda^\learner_{\vc{x}}},
\end{align*}
where $h(u) = 2\frac{(1+u)\ln{(1+u)}-u}{u^2}$ (Lemma \ref{lem: poisson tail}). 
\end{proof}

\subsection{Proof of Lemma \ref{lem: estimation guarantee}}\label{proof: estimation guarantee}
\begin{proof}
Our final estimator of the partial derivative is given by
\begin{equation*}
    \widehat{\frac{\partial U}{\partial\lambda^\learner_{\vc{x}}} }  \equiv T \sum_{n=0}^{n^{\prime}}\widehat{\Delta^\learner_{\vc{x}}(\vc{\lambda}^\learner,n) }\mathrm{Pr}[n^\learner_{\vc{x}} = n],\notag 
\end{equation*}
where
$$
    \widehat{\Delta^\learner_{\vc{x}}(\vc{\lambda}^\learner,n)} = \frac{1}{\nsamp}\sum_{j=1}^{\nsamp}(G^\learner(\vc{n}^{\learner,j}|_{n_{\vc{x}}^{\learner,j} = n+1}) - G^\learner(\vc{n}^{\learner,j}|_{n_{\vc{x}}^{\learner,j} = n})).\notag
$$
We define 
\begin{align*}
\textstyle X^j(n) = \frac{G^\learner(\vc{n}^{\learner,j}|_{n^{\learner,j}_{\vc{x}} = n+1}) - G^\learner(\vc{n}^{\learner,j}|_{n_{\vc{x}}^{\learner,j} = n}) - \Delta^\learner_{\vc{x}}(\vc{\lambda}^\learner,t)}{G_\mathrm{MAX}},
\end{align*}
where $$G_\mathrm{MAX} = \max_{\learner\in\learners, \vc{x}\in\xset}(G^\learner(\vc{e}_{\vc{x}})-G^\learner(\vc{0})).$$ We have $|X^j(n)|\leq 1$, because \begin{align*}G_\mathrm{MAX} &\geq G^\learner(\vc{e}_{\vc{x}})-G^\learner(\vc{0}) \\&\geq G^\learner(\vc{n}^\learner|n_{\vc{x}} = n+1)-G^\learner(\vc{n}^\learner|n_{\vc{x}} = n),\end{align*} for any $\learner\in\learners, \vc{x}\in\xset$, $n\geq 0$. By Chernoff bounds described by Theorem A.1.16 in \cite{alon2004probabilistic}, we have $$\mathrm{Pr}\left[\left|\sum_{j=1}^{N}\sum_{n = 0}^{n = n^{\prime}}X^j(n)\right|> c\right]\leq 2e^{-c^2/2N(n^{'}+1)}.$$ 
Suppose we let $c = \delta \cdot N/T$, where $\delta$ is the step size, then we have 
\begin{align*}
&|\widehat{\frac{\partial U}{\partial\lambda^\learner_{\vc{x}}} } - \mathrm{HEAD}^\learner_{\vc{x}}|   \\
\leq & \left|\sum_{n=0}^{n^{\prime}}\sum_{j=1}^{N} \textstyle \frac{(G^\learner(\vc{n}^{\learner,j}|_{n^{\learner,j}_{\vc{x}} = n+1}) - G^\learner(\vc{n}^{\learner,j}|_{n_{\vc{x}}^{\learner,j} = n}) - \Delta^\learner_{\vc{x}}(\vc{\lambda}^\learner,n))}{N} T\right| \\
= & \textstyle \left|\sum_{t=0}^{n^{\prime}}\sum_{j=1}^{N}X^j(n)\right|\cdot \frac{T}{N}\cdot G_\mathrm{MAX}\notag \leq \delta\cdot G_\mathrm{MAX},
\end{align*}
with probability greater than $1 - 2\cdot e^{-\delta^2N/2T^2(n^{\prime}+1)}$. By Lemma~\ref{lem: HEAD bound}, we have $\mathrm{HEAD}^\learner_{\vc{x}}\geq (1-\mathrm{P}[n^\learner_{\vc{x}}\geq n^{\prime} + 1])\cdot \frac{\partial U}{\partial \lambda^\learner_{\vc{x}}}$. Thus, we have
\begin{equation}\label{eq: estimation quality}
      \textstyle - \delta\cdot G_\mathrm{MAX}\leq\frac{\partial U}{\partial \lambda^\learner_{\vc{x}}} - \widehat{\frac{\partial U}{\partial\lambda^\learner_{\vc{x}}} }\leq \delta\cdot G_\mathrm{MAX} + \mathrm{Pr}[n^\learner_{\vc{x}}\geq n^{\prime} + 1]\cdot\frac{\partial U}{\partial \lambda^\learner_{\vc{x}}}.
\end{equation}
We now use the superscript $k$ to represent the parameters for the $k$th iteration: we find $\vc{v}^k\in\mathcal{D}$ that maximizes $\langle \vc{v}^k,\widehat{\nabla U(\vc{\lambda}^k)}\rangle$. 
Let $\vc{u}^{k}\in\mathcal{D}$ be the vector that maximizes $\langle \vc{u}^k,\nabla U(\vc{\lambda}^k)\rangle$ instead and define $\mathrm{P_{MAX}} = \max_{k=1,\dots, K}\mathrm{P}^k_\mathrm{MAX}$ where $\mathrm{P}^k_\mathrm{MAX} = \max_{l\in\learners,\x\in\xset} \mathrm{P}[n^{\learner,k}_{\x}\geq t^{\prime} + 1]$ and $\lambda_\mathrm{MAX} \equiv \max_{\vc{\lambda}\in\mathcal{D}}\sum_{\learner\in\learners}||\vc{\lambda}^{\learner}||_1$. We have
\begin{align*}
    &\langle \vc{v}^k,\nabla U(\vc{\lambda}^k)\rangle 
    \geq \langle \vc{v}^k,\widehat{\nabla U(\vc{\lambda}^k)}\rangle - \lambda_{\text{MAX}}\delta\cdot G_\mathrm{MAX}\\
    \geq & \langle \vc{u}^{k},\widehat{\nabla U(\vc{\lambda}^k)}\rangle - \lambda_{\text{MAX}}\delta\cdot G_\mathrm{MAX}\\
    \geq & (1-\mathrm{P_{MAX}})\cdot\langle \vc{u}^{k},\nabla U(\vc{\lambda}^k)\rangle - 2\lambda_{\text{MAX}}\delta\cdot G_\mathrm{MAX},
\end{align*}
where the first and last inequalities are due to \eqref{eq: estimation quality} and the second inequality is because $\vc{v}^k$ maximizes $\langle \vc{v}^k,\widehat{\nabla U(\vc{\lambda}^k)}\rangle$. The above inequality requires the satisfaction of \eqref{eq: estimation quality} for every partial derivative. By union bound, the above inequality satisfies with probability greater than $1 - 2|\xset||\mathcal{L}|\cdot e^{-\delta^2N/2T^2(n^{\prime}+1)}$.
\end{proof}

\end{document}